\documentclass[a4paper,fleqn,usenatbib]{mnras}
\newcommand \bibPath{./}
\newcommand \figPath{./}
\newcommand \redColor{}
\newcommand \blueColor{}
\renewcommand\labelenumi{(\roman{enumi})}
\renewcommand\theenumi\labelenumi

\usepackage{newtxtext,newtxmath}
\usepackage[T1]{fontenc}
\usepackage{ae,aecompl}

\usepackage{graphicx}	% Including figure files
\usepackage{subfigure}
\usepackage{amsmath}	% Advanced maths commands
\usepackage{amssymb}	% Extra maths symbols

\usepackage{pdflscape}	% Landscape pages
\usepackage{listings}
\usepackage{multirow}
\usepackage{multicol}  % Multi-column entries in tables

\usepackage{textcomp}
\usepackage{epstopdf}
\usepackage{natbib}
\usepackage{breakurl}
\hypersetup{breaklinks}

\usepackage{etoolbox}
\makeatletter
\patchcmd\@combinedblfloats{\box\@outputbox}{\unvbox\@outputbox}{}{%
   \errmessage{\noexpand\@combinedblfloats could not be patched}%
}%
 \makeatother

\title[FPFS Shear Estimator: Performance on Image Simulations ]{Fourier Power Function Shapelets (FPFS) Shear Estimator: Performance on Image Simulations}
\author[Li et al.]{Xiangchong Li,$^{1,2}$\thanks{E-mail: xiangchong.li@ipmu.jp}
Nobuhiko Katayama,$^{2}$
Masamune Oguri,$^{1,2,3}$
Surhud More$^{2,4}$
\\
$^{1}$Department of Physics, University of Tokyo, Tokyo 113-0033, Japan,\\
$^{2}$Kavli Institute for the Physics and Mathematics of the Universe (Kavli IPMU),\\
UTIAS, Tokyo Institutes for Advanced Study, University of Tokyo, Chiba 277-8583, Japan,\\
$^{3}$Research Center for the Early Universe, University of Tokyo, Tokyo 113-0033, Japan,\\
$^{4}$The Inter-University Center for Astronomy and Astrophysics (IUCAA),Post Bag 4, Ganeshkhind, Pune 411007, India.
}

\begin{document}
\label{firstpage}
\pagerange{\pageref{firstpage}--\pageref{lastpage}}
\maketitle

\begin{abstract}
We reinterpret the shear estimator developed by \citet{Z11} within the framework of Shapelets and propose the Fourier Power Function Shapelets (FPFS) shear estimator. Four shapelet modes are calculated from the power function of every galaxy's Fourier transform after deconvolving the Point Spread Function (PSF) in Fourier space. We propose a novel normalization scheme to construct dimensionless ellipticity and its corresponding shear responsivity using these shapelet modes. Shear is measured in a conventional way by averaging the ellipticities and responsivities over a large ensemble of galaxies. With the introduction and tuning of a weighting parameter, noise bias is reduced below one percent of the shear signal. We also provide an iterative method to reduce selection bias. The FPFS estimator is developed without any assumption on galaxy morphology, nor any approximation for PSF correction. Moreover, our method does not rely on heavy image manipulations nor complicated statistical procedures. We test the FPFS shear estimator using several HSC-like image simulations and the main results are listed as follows. (i) For simulations which only contain isolated galaxies, the amplitude of multiplicative bias is below $1\times 10^{-2}$. (ii) For more realistic simulations which also contain blended galaxies, the blended galaxies are deblended by the first generation HSC deblender before shear measurement. Multiplicative bias of $(-5.71\pm 0.31) \times 10^{-2}$ is found. The blending bias is calibrated by image simulations. Finally, we test the consistency and stability of this calibration.
\end{abstract}


\section{Introduction}
Light from background galaxies is deflected by the inhomogeneous foreground density distributions along the line-of-sight.  As a consequence, the images of background galaxies are slightly but coherently distorted. Such phenomenon is generally known as weak lensing. Weak lensing imprints the information of the foreground density distribution to the background galaxy images along the line-of-sight \citep{bookdodelson2017}. There are two types of weak lensing distortions, namely magnification and shear. Magnification isotropically changes the sizes and fluxes of the background galaxy images. On the other hand, shear anisotropically stretches the background galaxy images. Magnification is difficult to observe since it requires prior information about the intrinsic size (flux) distribution of the background galaxies before the weak lensing distortions \citep{MagZhang05}. In contrast, with the premise that the intrinsic background galaxies have isotropic orientations, shear can be statistically inferred by measuring the coherent anisotropies from the background galaxy images. Therefore, weak lensing offers a direct probe into the large scale structure of the Universe \citep[see][for recent reviews]{revKilbinger15,revRachel17} and it becomes the main target of several ongoing and upcoming surveys, including the Kilo-Degree Survey\footnote{\url{http://kids.strw.leidenuniv.nl/index.php}} \citep{KIDS13}, the Subaru Hyper Suprime-Cam (HSC) survey\footnote{\url{http://hsc.mtk.nao.ac.jp/ssp/}} \citep{HSC1-data}, the Dark Energy Survey\footnote{\url{http://www.darkenergysurvey.org/}} \citep{DES16}, the Large Synoptic Survey Telescope\footnote{\url{http://www.lsst.org/}} \citep{LSSTScienceBook}, the Euclid satellite mission\footnote{\url{http://sci.esa.int/euclid/}} \citep{Euclid2011}, and the WFIRST satellite mission\footnote{\url{http://wfirst.gsfc.nasa.gov/}} \citep{WFIRST15}.

Accurate shear measurement from galaxy images is challenging for the following reasons.
Firstly, galaxy images are smeared by Point Spread Functions (PSFs) as a result of diffraction by telescopes and the atmosphere, which is generally known as PSF bias.
Secondly, galaxy images are contaminated by background noise and Poisson noise originating from the particle nature of light, which is generally known as noise bias.
Thirdly, the complexity of galaxy morphology makes it difficult to fit galaxy shapes within a parametric model, which is generally known as model bias.
Fourthly, galaxies are heavily blended for deep surveys such as the HSC survey \citep{HSC1-pipeline}, which is generally known as blending bias.
Finally, selection bias emerges if the selection procedure does not align with the premise that intrinsic galaxies are isotropically orientated, which is generally known as selection bias.

Traditionally, several methods have been proposed to estimate shear from a large ensemble of smeared, noisy galaxy images. These methods is classified into two categories. The first category includes moments methods which measure moments weighted by Gaussian functions from both galaxy images and PSF models. Moments of galaxy images are used to construct the shear estimator and moments of PSF models are used to correct the PSF effect \citep[e.g.,][]{KSB95,GB02,Regaussianization}.
The second category includes fitting methods which convolve parametric Sersic models \citep{Sersic1963} with PSF models to find the parameters which best fit the observed galaxies. Shear is subsequently determined from these parameters \citep[e.g.,][]{LENSFIT1,im3Shape}.
Unfortunately, these traditional methods suffer from either model bias \citep{morphologyBiasGB10} originating from assumptions on galaxy morphology, or noise bias \citep[e.g.,][]{noiseBiasRefregier2012,noiseBiasOkura2018} due to nonlinearities in the shear estimators.

In contrast, \citet[ZK11]{Z11} measures shear on the Fourier power function of galaxies. ZK11 directly deconvolves the Fourier power function of PSF from the Fourier power function of galaxy in Fourier space. Moments weighted by isotropic Gaussian kernel\footnote{The Gaussian kernel is termed target PSF in the original paper of ZK11} are subsequently measured from the deconvolved Fourier power function. Benefiting from the direct deconvolution, the shear estimator of ZK11 is constructed with a finite number of moments of each galaxies. Therefore, ZK11 is not influenced by both PSF bias and model bias. 

{\redColor
We take these advantages of ZK11 and reinterpret the moments defined in ZK11 as combinations of shapelet modes. Shapelets refer to a group of orthogonal functions which can be used to measure small distortions on astronomical images \citep{ShapeletsI}. Based on this reinterpretation, we propose a novel normalization scheme to construct dimensionless ellipticity and its corresponding shear responsivity using four shapelet modes measured from every galaxies. Shear is measured in a conventional way by averaging the normalized ellipticities and responsivities over a large ensemble of galaxies. However, such normalization scheme introduces noise bias due to the nonlinear forms of the ellipticity and responsivity. With the introduction and tunning of a weighting parameter, the noise bias is reduced below one percent of shear signal. Furthermore, with the introduction of FPFS flux, we provide a iterative method to reduce selection bias below one percent of shear signal. This novel shear estimator is termed Fourier Power Function Shapelets (FPFS) shear estimator.

We test the performance of our FPFS shear estimator with several HSC-like image simulations. The tests on isolated galaxies show that the amplitude of multiplicative bias is below $1 \times 10^{-2}$ with some tuning of the weighting parameters. We combine our method with the first generation HSC deblender \citep{HSC1-pipeline} to measure shear from blended galaxies. Multiplicative bias of $(-5.71\pm 0.31) \times 10^{-2}$ is subsequently found for samples which also contain blended galaxies. The blending bias is calibrated using the realistic galaxy sample of the HSC-like simulations and finally the consistency of this calibration is checked.
}

Recently, several new methods have been proposed to reduce the multiplicative bias to a few parts in a thousand for isolated galaxies.
\citet{Z17} shows the latest development of ZK11 which is generally known as FOURIERQUAD (FQ hereafter). FQ determines two components of shear by re-symmetrizing the PDF of two spin-2 moments which are measured from the power function of galaxy's Fourier transform. This re-symmetry method only uses linear observable measured from Fourier power function to construct shear estimator so it is not influenced by noise bias. However, FQ has not provided solution to selection bias.
{\blueColor Bayesian Fourier Domain \citep[BFD hereafter;][]{BFD14,BFD16} uses Bayesian formalism to measure shear from the full Bayesian posterior so the formalism is not influenced by noise bias. BFD is the first method which provides solution to selection bias. BFD requires noiseless distribution of galaxy population over parameter space as a prior which should be constructed from deep exposures.}
Metacalibration \citep{metacal1,metacal2} proposes to find the shear responsivity for ellipticity defined by any algorithm through adding artificial shear to each observed galaxy. Shear can be inferred by averaging over ellpticities and responsitivities of a large ensemble of galaxies. {\redColor  Metacalibration adds inversely sheared noise image to galaxy images to remove noise bias and it also provides solution to selection bias. Several galaxy image simulations using realistic galaxy images \citep{metacal2} have proved that the multiplicative bias, including both noise bias and selection bias, for metacalibration is below $1\times 10^{-3}$. Moreover, metacalibration has been successfully applied to DES survey \citep{DEScat17}. Therefore, metacalibration is believed to be the most promising shear estimator in the weak lensing community. 

Comparing with these unconventional methods, the FPFS estimator does not involve complicated statistical procedure and it does not need prior information from deep exposures. Furthermore, FPFS shear estimator does not rely on heavy image manipulations. {\blueColor Since only four shapelet modes are required to construct the FPFS shear estimator, our algorithm is computationally fast.} Our FPFS method is an independent method that adopts different approach and assumptions from these unconventional methods. Although we always check the accuracy using image simulations, it is possible we might get biased results in real situations due to effects that were not included in the image simulations, but such biases might be unnoticed if we have only one method for the shear measurement. So comparisons using different methods are always very important.
}

This paper is organized as follows. Section 2 explores the analytical derivation of the FPFS formalism. Section 3 tests and calibrates the newly developed FPFS method using the HSC-like simulations. Section 4 checks the consistency and stability of the calibrated estimator. Section 5 provides a summary and outlook.

\section{Method}
\label{sec:Method}

This section is organized as follows. Section \ref{sub:powerSpectrum} explains why the shear estimator is constructed on the Fourier power function of galaxy images. Section \ref{sub:Shapelets} derives the shear estimator without consideration of photon noises. Section \ref{sub:noiBias} provides a solution to the noise bias. Section \ref{sub:selectMeth} discusses the selection bias.

\begin{figure*}%[!ht]
\includegraphics[width=0.8\textwidth]{\figPath 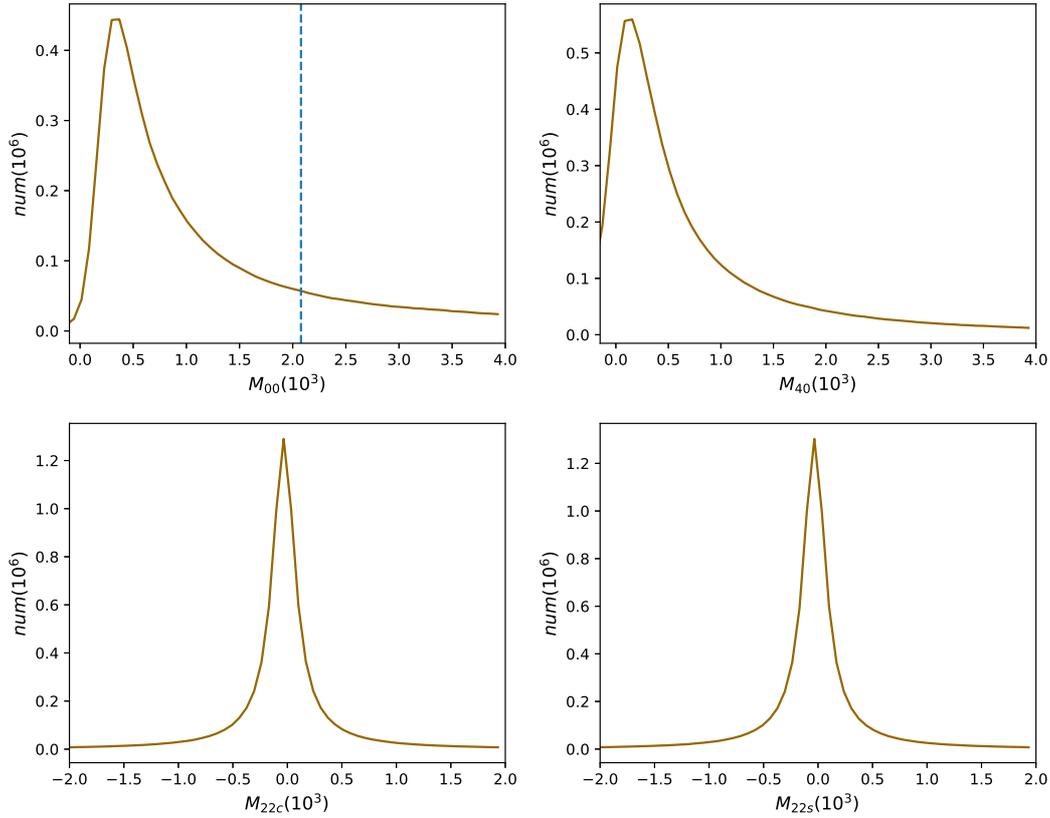}
\caption{The solid lines show the histograms of the shapelet modes $M_{nm}$. These observables are measured from the modelled galaxy sample of the HSC-like simulations. The aperture ratio and shapelets scale ratio are kept to ($\alpha=4$,$\beta=0.85$). The dashed line in the upper lefter panel shows the value of $\Delta$ used to normalize weighting parameter $C$ as shown in eq. (\ref{nu_define}).}\label{fig_MnmHist}
\end{figure*}
\subsection{Fourier power function}
\label{sub:powerSpectrum}
Weak lensing signal is presumed to be constant on the galaxy scale. For a single galaxy, the lensing effect can be expressed as a mapping of the galaxy surface brightness field from the intrinsic plane $\vec{x}'$ to the lensed plane $\vec{x}$, 
\begin{equation}\label{shear_distortion}
\begin{split}
\vec{x}&=\mathbf{S} \vec{x}',\\
 f(\vec{x})&=\bar{f}(\vec{x}'),
\end{split}
\end{equation}
where $\bar{f}$ and $f$ are the surface brightness distributions of the intrinsic and the lensed galaxy, respectively. $\mathbf{S}$ is the Jacobian matrix which is defined as
\begin{equation}\label{jacobian_matrix}
\mathbf{S}= (1+\kappa)\begin{pmatrix}
1+g_1&g_2\\
g_2  &1-g_1
\end{pmatrix}.
\end{equation}
The two components of the reduced shear ($g_1,g_2$) cause the anisotropic stretching of the galaxy image and the convergence $\kappa$ describes a change in galaxy size and brightness. The PSF effect can be expressed as a convolution between the lensed galaxy ($f$) and the PSF ($g$):
\begin{equation}\label{PSF convolution}
f_o(\vec{x}_o)=\int{g(\vec{x}_o-\vec{x})f(\vec{x})d^2 x}.
\end{equation}

The Fourier power function of the galaxy is defined as
\begin{align*}
\tilde{f}_o(\vec{k})&=\int f_o(\vec{x}_o) e^{-i\vec{k}\vec{x}_o} d^2x_o,\\
\tilde{F}_o(\vec{k})&=|\tilde{f}_o(\vec{k})|^2,
\end{align*}
where $\tilde{f}_o(\vec{k})$ and $\tilde{F}_o(\vec{k})$ are the Fourier transform and the Fourier power function of the galaxy image, respectively.
In order to ensure the stability of Fourier transform in real observations, it is necessary to define a boundary for each galaxy and subsequently mask the pixels outside the boundary with zero.
\citet[LZ17]{Li17Auto} proposed to use a top-hat aperture around the galaxy center to define its boundary. The top-hat filter is defined as
\begin{equation}\label{top-hat definition}
T(\vec{x})=
\begin{cases}
1,& |\vec{x}-\vec{x}_c|<r_{ \text{cut}}\\
0,& |\vec{x}-\vec{x}_c|\geq r_{\text{cut}}
\end{cases},
\end{equation}
where $\vec{x}_c$ is the galaxy centroid and $r_{\rm cut}$ is the aperture radius. The ratio between the aperture radius ($r_{\rm{cut}}$) and trace radius of galaxy ($r_{g}$) is termed the aperture ratio, which is defined as 
\begin{align*}
\alpha=\frac{r_{\rm cut}}{r_{g}},
\end{align*}
where the trace radius is determined by the galaxy's quadrupole moments matrix measured by the re-Gaussianization method \citep{Regaussianization}. Denoting the quadrupole moments matrix as $\textbf{Q}$, the trace radius is defined as
\begin{equation}
r_{g}=\sqrt{\frac{tr(\textbf{Q})}{2}}.
\end{equation}
To avoid steep cut on the galaxy's light profile, the aperture ratio should not be too small. Neither should the aperture ratio be too big, otherwise the measurement can be easily influenced by light from neighbouring objects. The influence of $\alpha$ on the accuracy of the shear estimator is discussed in Section \ref{sub:cutRad}.

One reason for using the Fourier power function is that the centroid of the Fourier power function is well defined even in the presence of noise since the Fourier power function is always symmetric around its zero point where $\vec{k}=\vec{0}$ \citep{Z11}. Therefore, we do not need to worry about the bias caused by the off-centering when calculating high order shapelet modes.
Another reason is that the PSF effect can be removed by dividing the PSF Fourier power function ($\tilde{G}$) from the observed galaxy Fourier power function 
\begin{equation}\label{PSF deconvolution_fourier}
\tilde{F}(\vec{k})=\frac{\tilde{F}_o(\vec{k})}{\tilde{G}(\vec{k})}.
\end{equation}
The reader may concern that the PSF deconvolution in Fourier space amplify the noise at small scale. However, as shown in Section \ref{sub:Shapelets}, we project the deconvolved galaxy Fourier power function to the basis vectors of polar shapelet \citep{polar_Shapelets} to ensure the stability of the estimator.

\begin{figure*}%[t]
\begin{center}
\includegraphics[width=0.8\textwidth]{\figPath 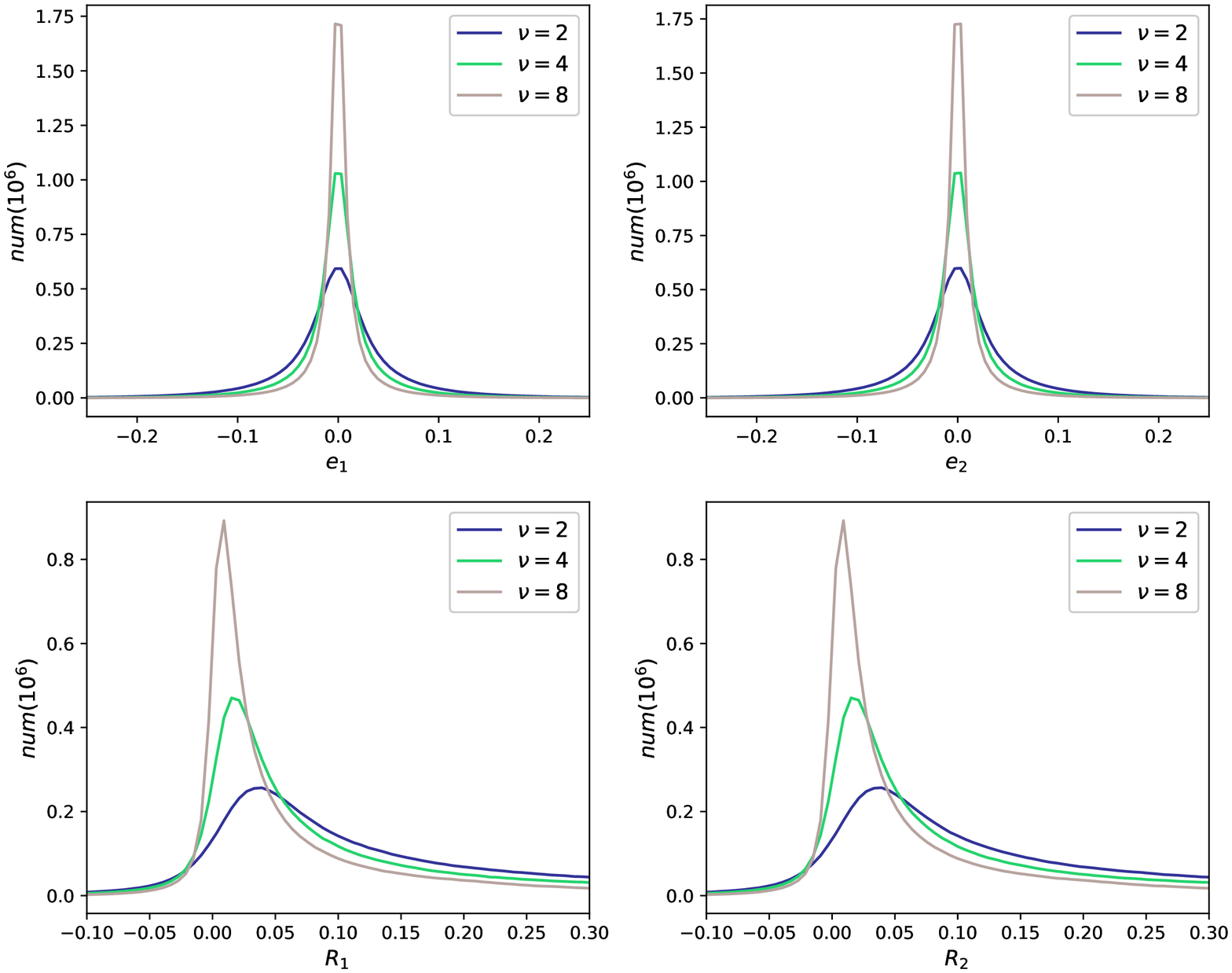}
\end{center}
\caption{The histograms of the FPFS ellipticity $e_{1,2}$ and responsivity $R_{1,2}$ for different values of $\nu$. The aperture ratio and shapelets scale ratio are kept to ($\alpha=4$,$\beta=0.85$). These observables are measured from the modelled galaxy sample of the HSC-like simulations. Lines with different colors are for different values of $\nu$, where $\nu$ is set to $2,4,8$.}\label{fig_ElliHist}
\end{figure*}

\subsection{Ellipticity and responsivity}
\label{sub:Shapelets}
\subsubsection{unnormalized estimator}
The Fourier power function of the intrinsic galaxy is distorted by {\blueColor ${\mathbf{S}}^{-1}$} due to the weak lensing effect. Therefore, the weak lensing signal can be inferred from the Fourier power function of the galaxy after PSF deconvolution shown in eq. (\ref{PSF deconvolution_fourier}). In order to measure the shear signal, we project the deconvolved galaxy Fourier power function onto four basis vectors of polar shapelets. The polar shapelet basis vectors \citep{polar_Shapelets} are generally defined as
\begin{align*}
\chi_{nm}(r,\theta)&=\frac{(-1)^{(n-|m|)/2}}{\sigma^{|m|+1}}\left\lbrace\frac{[(n-|m|)/2]!}{\pi[(n+|m|)/2]!}\right\rbrace^\frac{1}{2}\\
&\times r^{|m|}L^{|m|}_{\frac{n-|m|}{2}}\left(\frac{r^2}{\sigma^2}\right)e^{-r^2/2\sigma^2} e^{-im\theta},
\end{align*}
where $L^{p}_{q}$ are the Laguerre Polynomials, $n$ is the radial number and $m$ is the spin number, $\sigma$ determines the scale of shapelet functions. The ratio between $\sigma$ and the scale radius of PSF Fourier power function ($r_{\text{pp}}$)  is denoted as $\beta$, where
\begin{equation}
\beta=\frac{\sigma}{r_{\text{pp}}},
\end{equation}
and $\beta$ is termed shapelets scale ratio. $r_{\text{pp}}$ is measured from {\redColor the noiseless PSF model} in the same way as LZ17. After recording the maximum value of the Fourier power function of {\redColor the noiseless PSF model}, the area of pixels (where the value is greater than $e^{-0.5}$ of the recorded maximum value) is measured and denoted as $A$. The scale radius of PSF Fourier power function ($r_{\text{pp}}$) is calculated as
\begin{align*}
r_{\text{pp}}=\sqrt{\frac{A}{\pi}}.
\end{align*}
The projection of the deconvolved galaxy converges only if the shapelet basis vectors are more compact than PSF in Fourier apace, namely $\beta$ is required to be smaller than one. It is worth noting that $\alpha$ and $\beta$ determine the scales we focus on in real space and Fourier space, respectively. Section \ref{sub:cutRad} demonstrates the accuracy of the FPFS shear estimator with different choices of $\alpha$ and $\beta$ using the modelled galaxy sample of the HSC-like simulations.

The projection factors, which is termed shapelet modes, are denoted as $M_{nm}$, where
\begin{align}\label{Shapelets_decompose}
M_{nm}=\int \chi_{nm}^{*} \tilde{F}(r,\theta) r dr d\theta.
\end{align}
$M_{nmc}$ and $M_{nms}$ are used to denote the real and imaginary part of $M_{nm}$ when $m>0$. {\redColor Due to the symmetry of power function, all of the odd order shapelet moments vanish.} Four shapelet modes are used to construct the FPFS ellipticity and the histograms of these modes measured from the modelled galaxy sample of the HSC-like simulations are shown in Fig. \ref{fig_MnmHist}. In order to quantify the spread of $M_{00}$, the value of $M_{00}$ at which its histogram drops below $1/8$ of its maximum {\blueColor(on the side of higher $M_{00}$)} is denoted as $\Delta$. 

The transformation formulas of the shapelet modes under the influence of shear have been given by \citet{polar_Shapelets}, which are
\begin{equation}\label{Shapelets_Moments_shear_Transform}
\begin{split}
M_{22c}&=\bar{M}_{22c}-\frac{\sqrt{2}}{2}g_1(\bar{M}_{00}-\bar{M}_{40})\\
&+\sqrt{3}g_1 \bar{M}_{44c}+\sqrt{3} g_2 \bar{M}_{44s},\\
M_{22s}&=\bar{M}_{22s}-\frac{\sqrt{2}}{2}g_2(\bar{M}_{00}-\bar{M}_{40})\\
&-\sqrt{3}g_2 \bar{M}_{44c}+\sqrt{3} g_1 \bar{M}_{44s},\\
M_{00} &=\bar{M}_{00}+\sqrt{2}(g_1\bar{M}_{22c}+g_2\bar{M}_{22s}),\\
M_{40} &=\bar{M}_{40}-\sqrt{2}(g_1\bar{M}_{22c}+g_2\bar{M}_{22s})\\
&+2\sqrt{3}(g_1\bar{M}_{62c}+g_2\bar{M}_{62s}),
\end{split}
\end{equation}
where $\bar{M}_{nm}$ represent the intrinsic shapelet modes and $M_{nm}$ represent the sheared shapelet modes.
Shear can be inferred by taking the expectation values on the both sides of eq. (\ref{Shapelets_Moments_shear_Transform}). Assuming the galaxy ensemble is randomly selected without preference on any specific direction, the intrinsic spin-2 and spin-4 shapelet modes on the right hand side of eq. (\ref{Shapelets_Moments_shear_Transform}) reduce to zero. The population variance of these intrinsic spin-2 and spin-4 quantities causes the shape noise in shear estimation. Finally, we derive the shear estimator 
\begin{align}\label{unnormalized_estimator}
g_1=-\frac{\left\langle\sqrt{2}M_{22c}\right\rangle}{\left\langle M_{00}-M_{40}\right\rangle},\qquad
g_2=-\frac{\left\langle\sqrt{2}M_{22s}\right\rangle}{\left\langle M_{00}-M_{40}\right\rangle},
\end{align}
which is mathematically equivalent to the shear estimator proposed by \citet[see eq. (42)]{Z11}. However, such shear estimator is dominated by the shape noise since bright galaxies are overweighted in the shear estimation. A normalization scheme is required to re-weight the shapelet modes and reduce the shape noise of the shear estimation. 

\subsubsection{normalized estimator}
We introduce a novel normalization scheme using these four shapelet modes. Firstly, the dimensionless FPFS ellipticity is defined as
\begin{align}\label{ellipticity_define}
e_1=\frac{M_{22c}}{M_{00}+C},\qquad
e_2=\frac{M_{22s}}{M_{00}+C}.
\end{align}
The constant parameter $C$ is termed weighting parameter, which adjusts the relative weight between galaxies with different luminosities. We normalize the weighting parameter by $\Delta$ and denote the normalized weighting parameter as $\nu$, where
\begin{equation}\label{nu_define}
\nu=\frac{C}{\Delta}.
\end{equation}
With the definition of the responsivity
\begin{align}\label{response_define}
R_i=\frac{\sqrt{2}}{2}\frac{M_{00}-M_{40}}{M_{00}+C}+\sqrt{2}e_i^2,
\end{align}
the transformation of the FPFS ellipticity under the influence of shear is subsequently derived as
\begin{align}\label{ellipticity_transform}
e_1&= \bar{e}_1-g_1 \bar{R}_1+\sqrt{3}g_1\frac{\bar{M}_{44c}}{\bar{M}_{00}+C}+\sqrt{3}g_2\frac{\bar{M}_{44s}}{\bar{M}_{00}+C},\\\notag
e_2&= \bar{e}_2-g_2 \bar{R}_2-\sqrt{3}g_2\frac{\bar{M}_{44c}}{\bar{M}_{00}+C}+\sqrt{3}g_1\frac{\bar{M}_{44s}}{\bar{M}_{00}+C}.
\end{align}
\begin{figure*}%[!ht]
\begin{center}
\includegraphics[width=0.9\textwidth]{\figPath 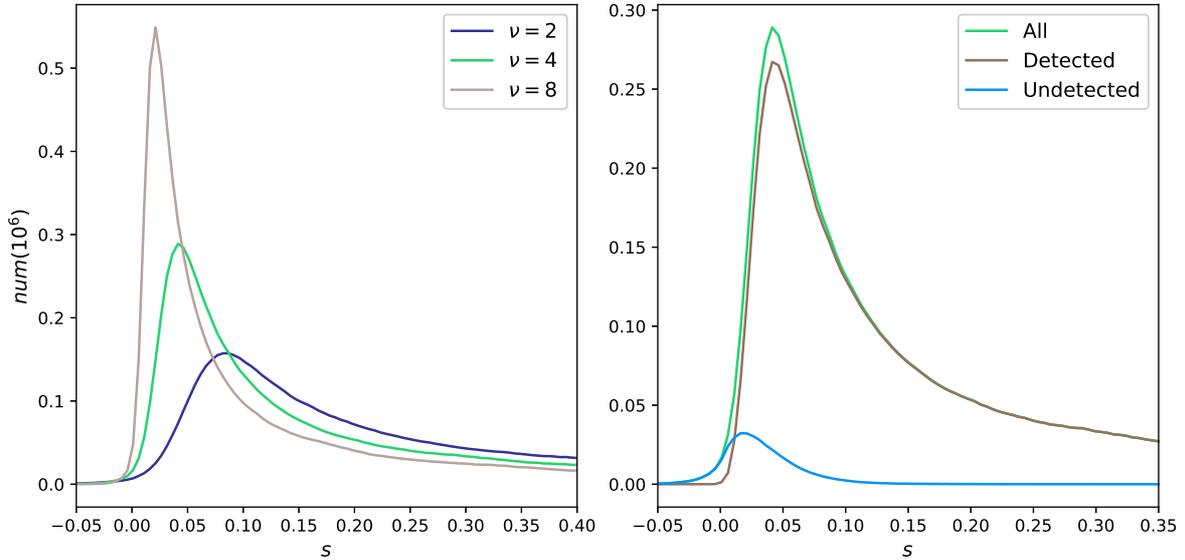}
\end{center}
\caption{The histograms of the FPFS fluxes with different setups of $\nu$. These observables are measured from the modelled galaxy sample of the HSC-like simulations. The aperture ratio and shapelets scale ratio are kept to ($\alpha=4$,$\beta=0.85$). (i) The left panel shows the histograms of the FPFS fluxes defined in eq. (\ref{select_define}). Lines with different colors correspond to different values of $\nu$. (ii) The right panel shows the histograms of detected galaxies, undetected galaxies and all of the galaxies. For the right panel we keep $\nu=4$.}\label{fig_selectHist}
\end{figure*}
The detailed derivation of eq. (\ref{ellipticity_transform}) for noiseless galaxies is shown in Appendix \ref{app:elliTrans}. Fig. \ref{fig_ElliHist} demonstrates the histograms of $e_{1,2}$ and $R_{1,2}$ for different choices of $\nu$. 

Given an ensemble of galaxies distorted by a constant shear, the shear signal is measured with the expectation of the ellipticities and the responsivities. If the galaxy ensemble is randomly selected, the galaxies within the ensemble have random orientations. Therefore, the expectation value of the spin-2 and spin-4 quantities in eq. (\ref{ellipticity_transform}) reduces to zero. {\blueColor The population variance of the intrinsic spin-2 and spin-4 quantities on the right hand side of eq. (\ref{ellipticity_transform}) causes the shape noise in shear estimation.}  Moreover, the expectation value of the intrinsic responsivities ($\langle \bar{R}_{1,2} \rangle$) are the same as the expectation value of the observed responsivities ($\langle R_{1,2} \rangle$). The shear estimator is consequently constructed as follows
\begin{align}\label{estimator_define}
g_i=-\left\langle e_i\right\rangle/\left\langle R_i\right\rangle.
\end{align}

In the averaging procedure of the shear estimation, the intrinsic responsivity ($\bar{R}_{1,2}$) acts as the weight on each galaxy. {\blueColor As shown in the left panel of Fig. \ref{fig_PrecisionNu}, $\nu$ changes the relative weight between galaxies with different luminosities. When $\nu$ increases, more weight is added to bright galaxies and, when $\nu$ goes to infinity, the shear estimator reduces to eq. (\ref{unnormalized_estimator}). Therefore, the precision of the shear estimation is dependent on the value of $\nu$, which is discussed in Section \ref{sub:cutRad}.}

In the absence of photon noises, $\nu$ can be set to any positive value in the shear estimation. However, in the presence of photon noises, the accuracy of the shear estimation is also dependent on the value of $\nu$ as discussed in Section \ref{sub:noiBias} and Section \ref{sub:cutRad}.
\subsection{noise bias}
\label{sub:noiBias}
Photon noises include Poisson noise and background noise. Although the amplitude of Poisson noise correlates with the surface brightness distributions of galaxies, its phase does not. Under the further assumption that the background noise does not correlate with the surface brightness distributions of galaxies, the total noises remain uncorrelated with the surface brightness distributions of galaxies, even after the coadding process. Based on this premise, the averaged contamination of noise can be removed by subtracting the Fourier power function of noise from the galaxy Fourier power function \citep{Z15}. 

We propose to reconstruct the Fourier power function of noise using the noise correlation function measured from blank pixels. The details for the reconstruction of noise Fourier power function are shown in Appendix \ref{app:noiPow}.
{\redColor 
We caution that Poisson noise could make the noise correlation function on blank pixels (without detected sources) different from the noise correlation function on bright pixels (with detected sources) since the amplitude of Poisson noise is correlated with the readout of the pixel. On single exposures, Poisson noise is not correlated across pixels \citep{Z15} so the power function of Poisson noise can be estimated from the large wave number in Fourier space and subsequently subtracted from the Fourier power function of galaxies \citep{Z15}. The coadding process correlates Poisson noise across pixels and the correlation of Poisson noise is hard to estimate and remove. Since Poisson noise is proportional to the readout of the pixel, it does not cause severe problems on faint galaxies and we expect that the contribution of Poisson noise on bright galaxies are relatively small due to the high signal to noise ratio of bright galaxies. In this paper, we assume that the correlation function of noise is universal across each exposure and independent of the pixel readout.  The difference between the correlation function on blank pixels and that on bright pixels caused by Poisson noise will be discussed in details in our future work.}
\begin{figure*}%[!ht]
\begin{minipage}[t]{.48\textwidth}
\centering
\includegraphics[width=1.\textwidth]{\figPath 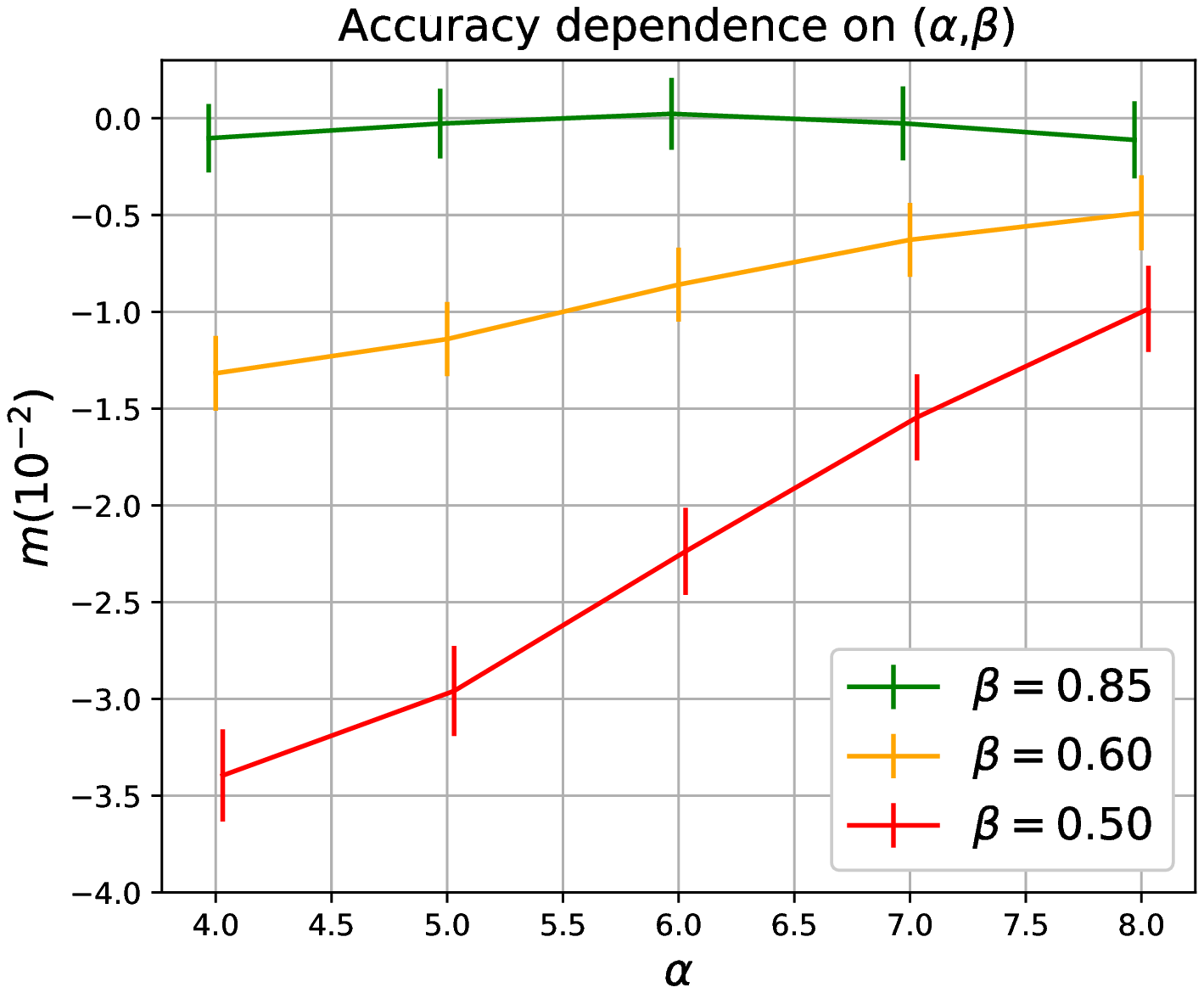}
\caption{This figure shows the accuracy of the FPFS shear estimator for different setups of $\alpha$ and $\beta$. The test is conducted on sample 2 of the GREAT3-HSC simulations. The weighting parameter $\nu$ is kept to $4$ for these results. The $x$-axis is the aperture ratio $\alpha$. The $y$-axis is the multiplicative bias. Lines with different color correspond to $\beta=0.50,0.60,0.85$.}
\label{fig_betaCut}
\end{minipage}\hfill
\begin{minipage}[t]{.48\textwidth}
\centering
\includegraphics[width=1.\textwidth]{\figPath 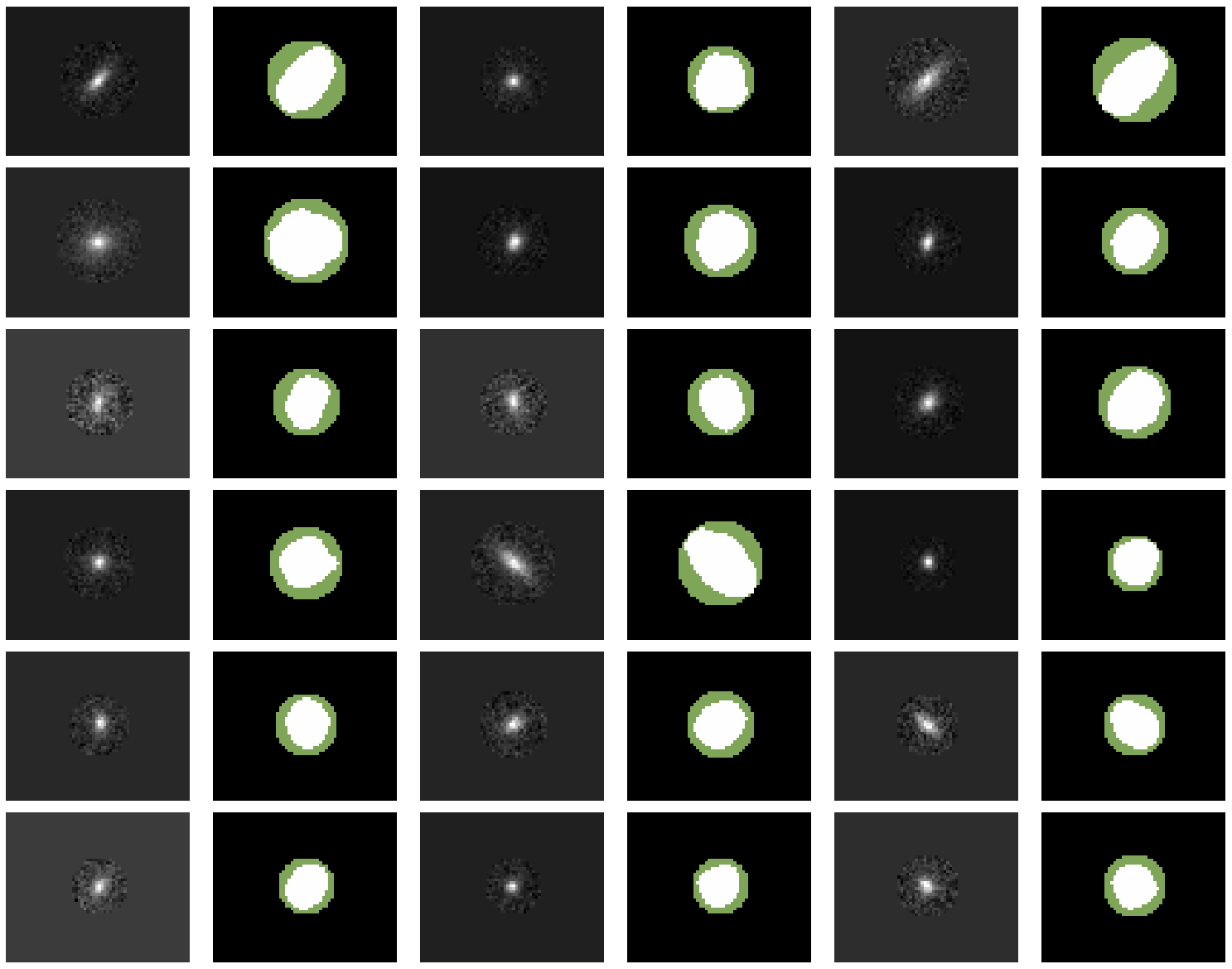}
\caption{The galaxy images and masks detected from sample 2 of the GREAT3-HSC simulations where the aperture ratio ($\alpha$) is set to $4$. White pixels on the masks show the detected footprints and the gray pixels represent the area within the aperture radius. }
\label{fig_galCut}
\end{minipage}
\end{figure*}

After the subtraction of noise power function, a zero-mean residual is left on the Fourier power function of the galaxy. The shapelet modes of such residual are denoted as $N_{nm}$. The expectation value of the shapelet modes of noise residual ($N_{nm}$) should be zero and they are assumed to be not correlated with the galaxy's shapelet modes ($M_{nm}$). When $\nu=0$, the residuals on the galaxy Fourier power functions cause bias to the shear estimation due to the nonlinear form of the FPFS ellipticity and responsivity. Taking $\left\langle e_1 \right\rangle$ as an example, the expectation value of $e_1$ changes to
\begin{align*}
\left\langle e_1\right\rangle&=\left\langle\frac{M_{22c}+N_{22c}}{M_{00}+N_{00}}\right\rangle\neq \left\langle\frac{M_{22c}}{M_{00}}\right\rangle,
\end{align*}
which does not equate the noiseless ellipticity. Therefore, the normalization procedure introduces the noise bias.
In order to reduce the noise bias originating from the nonlinearity of the ellipticity, we increase $\nu$ to make $M_{00}+C \gg N_{00}$. Subsequently, the expectation value of $e_1$ changes to
\begin{align}\label{noise_bias}
\left\langle e_1\right\rangle&=\left\langle \frac{M_{22c}}{M_{00}+C}(1+O(\epsilon^2))\right\rangle,
\end{align}
where 
\begin{align*}
\epsilon=\frac{N_{00}}{M_{00}+C}\ll 1.
\end{align*}
Thus the remaining noise bias is proportional to $\epsilon^2$. The expectation values of $e_2$ and $R_{1,2}$ have similar form as eq. (\ref{noise_bias}). On the other hand, the value of $\nu$ cannot be too large otherwise the bright galaxies are overweighted and the shear estimation is dominated by the shape noise. 
The dependence of the accuracy and precision on the value of $\nu$ is demonstrated in Section \ref{sub:cutRad} using the modelled galaxy sample of the HSC-like simulations.

In comparison, \citet{Z17} explores another statistical way to infer shear from large ensembles of galaxies which determines the shear by finding the value which best symmetrize the Probability Density Function (PDF) of the unnormalized spin-2 moments. Such re-symmetry method  equally attribute weight to different galaxies regardless of their brightness. \citet{Z17} proposed to use the unnormalized moments, where noises do not has nonlinear contribution, to avoid noise bias.

\subsection{selection bias}
\label{sub:selectMeth}

In this section, we discuss the selection bias caused by the improper selection of the galaxy ensemble. In order to define a selection procedure, one need to define a group of selection functions and their corresponding selection thresholds. A galaxy is counted as a member of the ensemble if all of its selection functions fall within the corresponding selection thresholds. Moreover, the selection functions should be isotropic (spin-0) quantities on the intrinsic plane to ensure that galaxies are isotropically selected. Galaxies should be isotropically selected since the shear measurement is based on the premise that the intrinsic galaxies within the galaxy ensemble have statistically isotropic orientations. 

To be more specific, we define the FPFS flux as
\begin{align}\label{select_define}
s = \frac{M_{00}}{M_{00}+C},
\end{align}
and use the FPFS flux as the selection function.
The left panel of Fig. \ref{fig_selectHist} shows the histograms of $s$ with different setups of $\nu$. In addition, the detection of galaxies is also a selection process which could cause bias to the shear measurement so we show the histogram of $s$ ($\nu=4$) for the undetected galaxies on the right panel of Fig. \ref{fig_selectHist}. It suggests that most of the undetected galaxies populate within the range $s<0.1$. 

The FPFS flux is also influenced by the shear and the relationship between the sheared FPFS flux ($s$) and the intrinsic FPFS flux ($\bar{s}$) is
\begin{align}\label{select_transform}
s = \bar{s}+\sqrt{2}g_1 \bar{e}_1(1-\bar{s})+\sqrt{2}g_2 \bar{e}_2(1-\bar{s}).
\end{align}
$\bar{s}$ is isotropic (spin-0) on the intrinsic plane but $s$ is not. Therefore, the selection using $s$ as the selection function is not an isotropic selection on the intrinsic plane. Such selection does not align with the premise that the intrinsic galaxies have isotropic orientations statistically and causes selection bias.
We provide an iterative method to reduce the selection bias as follows.
\begin{enumerate}
\item Estimate shear with selection $L< s < U$ and the estimated shear is denoted as $\hat{g}^A_{1,2}$.
\item Inversely transform the observed selection function $s$ into $s_{R}$ which is isotropic in the intrinsic plane with $\hat{g}^A_{1,2}$ according to eq. (\ref{select_transform}).
\item Re-estimate shear with selection $L<s_R<U$ and update the outcome of the shear measurement to $\hat{g}^B_{1,2}$.
\end{enumerate}

{\blueColor
The performance of this iterative method is demonstrated in Section \ref{sub:selectTest} using the HSC-like image simulations. The current Galaxy image simulations distorted a large ensemble of galaxies by a constant shear. Such setup of image simulations simplifies real observations since it does not consider the scatter of shear signal. The influence of the scatter of the shear to the selection bias revision is beyond the scope of this paper and it will be discussed in our future works.}  
\begin{figure}
\centering
\includegraphics[width=0.45\textwidth]{\figPath 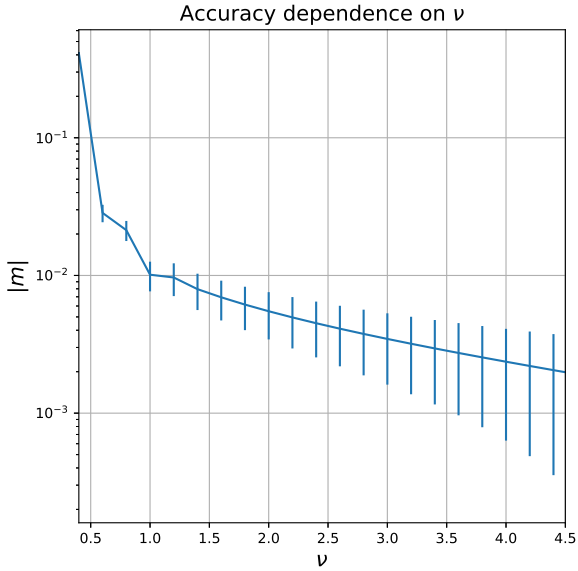}
\caption{This figure shows the accuracy of the FPFS shear estimator for different setups of $\nu$. The test is conducted on sample 2 of the GREAT3-HSC simulations. The aperture ratio and shapelets scale ratio are kept to ($\alpha=4$,$\beta=0.85$). The $x$-axis is the weighting parameter ($\nu$) and the $y$-axis is the absolute value of the multiplicative bias.}
\label{fig_AccuracyNu}
\end{figure}

\section{Test and Calibration}
\label{sec:Test}
\begin{figure*}%[!ht]
\begin{center}
\includegraphics[width=1.\textwidth]{\figPath 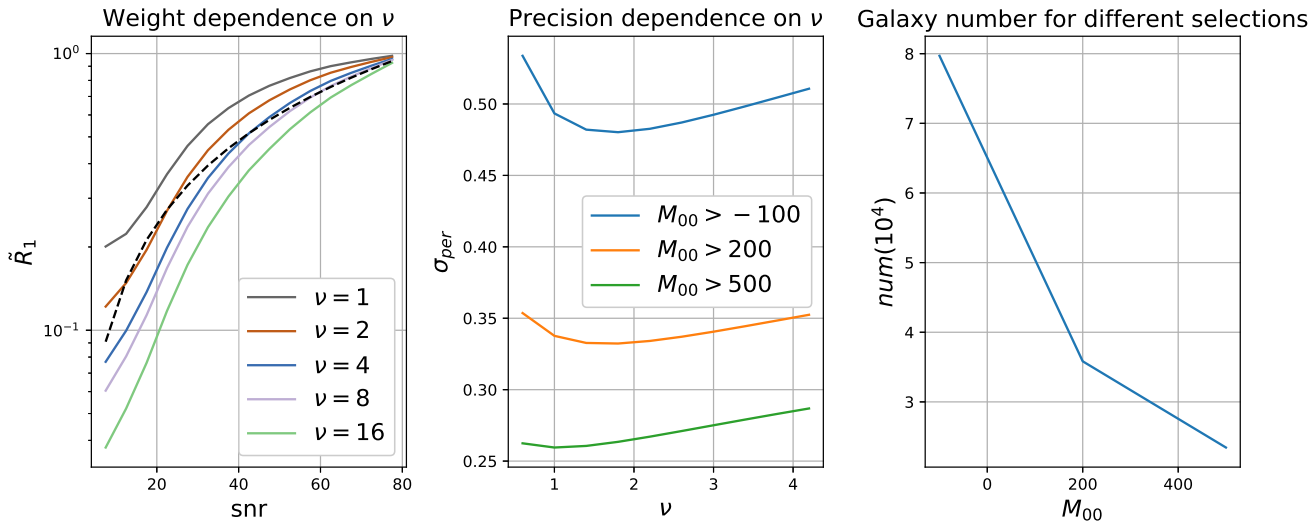}
\end{center}
\caption{The precision of the FPFS shear estimator for different setups of $\nu$. The aperture ratio and shapelets scale ratio are kept to ($\alpha=4$,$\beta=0.85$) for these tests. (i) The left panel shows that $\nu$ changes the relative weight between galaxies in different $S/N$ bins. The $x$-axis is the $S/N$ bins and the $y$-axis is the averaged $R_1$ normalized by the averaged $R_1$ of the bin at $75<S/N<80$. The solid lines with different colors are for different $\nu$. The dashed line demonstrates the case where weight ratio is proportional to the $S/N$ as a reference.  (ii) The middle panel shows the precision of the estimator for different setups of $\nu$ and different selections. The $x$-axis is the weighting parameter ($\nu$) and the $y$-axis is the shear measurement error per galaxy per component. Lines with different colors represent different lower selection threshold on $M_{00}$. (iii) The right panel shows the galaxy number for different lower selection threshold on $M_{00}$.}
\label{fig_PrecisionNu}
\end{figure*}
Testing shear estimators with galaxy image simulations is a long-standing tradition within the weak lensing community \citep[e.g.,][]{GREAT08,GREAT3}. Galaxy image simulations which match the real observational conditions can be used to calibrate shear measurements \citep{SHEARA}. To test and calibrate shear estimators, we use image simulations with large ensembles of galaxies distorted by a group of known input shears ($g_{1,2}$). Then, shears are measured from the simulated images by a specific method, where the estimated shear is denoted as $\hat{g}_{1,2}$. Based on the premise that the amplitude of the input shear is only a few percent, the estimated shear ($\hat{g}_{1,2}$) can be expressed as the first order Taylor expansion of the input shear ($g_{1,2}$)
\begin{align}\label{bias definition}
\hat{g}_{1,2}=(1+m_{1,2}) g_{1,2}+c_{1,2}.
\end{align}
Since all of the even order terms of the input shear are not correlated with the estimated shear, these even oder terms must be zero on average. Therefore, we only ignore third and higher odd order terms in eq. (\ref{bias definition}). $m_{1,2}$ and $c_{1,2}$ are termed multiplicative bias and additive bias, respectively. $m$ is used to denote the mean of the multiplicative bias over two shear components, namely $m=(m_1+m_2)/2$.

Generally, multiplicative biases depend on several properties, including galaxy shapes, galaxy luminosities, seeing conditions, noise properties and neighbouring objects. On the other hand, nonzero additive biases can be generated by some anisotropy in the PSF, anisotropic selection effects and the masks in observations. Biases should be modelled as multi-dimensional functions of a group of observed properties and the form of these functions are dependent on the estimator used for shear measurement. It is promising if we can find a shear estimator for which all of the biases are consistent with zero. Another solution is to model the biases with image simulations and calibrate a biased shear estimator on single galaxy level.

This section is organized as follows. Section \ref{sub:simulations} describes two sets of HSC-like galaxy image simulations used to test and calibrate our shear estimator. Section \ref{sub:pipeline} introduces the FPFS algorithm as a subroutine of the HSC pipeline \citep{HSC1-pipeline}. Section \ref{sub:cutRad} explores the dependence of the accuracy and precision of the FPFS shear estimator on the free parameters ($\alpha$ ,$\beta$ and $\nu$) using image simulations. Section \ref{sub:selectTest} and Section \ref{sub:neiObj} test the performance of our method on isolated galaxies and blended galaxies, respectively.
\subsection{Simulations}
\label{sub:simulations}
The HSC-like Bulge+Disk+Knot (BDK) simulation is an HSC version of the BDK simulation \citep{metacal2}. The simulation is generated by Galsim which is an open-source image simulation package \citep{GalSim}. We use Sersic models \citep{Sersic1963} which are fitted to the $25.2$ magnitude limited galaxy sample from the COSMOS data \footnote{\url{great3.jb.man.ac.uk/leaderboard/data/public/COSMOS_25.2_training_sample.tar.gz}} to simulate the bulge and disk of galaxies. The fluxes of these galaxies are scaled by a factor of $2.587$ to match the fluxes in HSC observation. In order to avoid repeating the exact parameters, we interpolate the joint radius-flux distribution by randomly rescaling the radius and flux of the original Sersic model. To simulate the knots of star formation, we distribute $N$ random points which statistically obey the Gaussian distribution around the center of the galaxies, where $N$ is a random number evenly distributed between $50$ and $100$. The ellipticity of the Gaussian distribution follows the ellipticity of Sersic model and the half light radius of the Gaussian distribution is fixed to $2.4$ pixels. The pixel scale of the simulation is set to the HSC pixel scale, namely $0.168''$. The fraction of the flux of the knots is a random number evenly distributed between $0\%$ and $10\%$. The galaxies are rotated to random directions and subsequently sheared by the same shear signal ($g_1=0.02,g2=0.00$). For the HSC-like BDK simulation, we use $g_1$ to determine the multiplicative bias and use $g_2$ to determine the additive bias. The galaxy images are convolved with a Moffat PSF \citep{Moffat1969}
\begin{equation}\label{Moffat PSF}
g_{m}(\vec{x})=[1+c(|\vec{x}|/r_p)^2]^{-\beta_m},  
\end{equation}
where $c=2^{\frac{1}{\beta_m-1}}-1$ is a constant parameter. The profile of the Moffat PSF is determined by $\beta_m$, where $\beta_{m}=3.5$. The scale of the Moffat PSF is determined by its Full Width Half Maximum (FWHM), where $\rm{FWHM}=0.6$.  The ellipticity of the Moffat PSF is set to $(e_1=0,e_2=0.025)$. Each convolved galaxy is placed around the center of a $64\times 64$ stamp. The HSC-like BDK simulation generate $4\times 10^8$ galaxies.

Galaxy images of the GREAT3-HSC simulations \citep{GREAT3HSC17} are also generated by GalSim using images from the COSMOS HST survey. The simulations provide four samples of galaxy images with different galaxy properties. The structure of simulations is described as follows. Each sample is divided into $800$ subfields; each subfield contains $10^4$ postage stamps; each postage stamp contains $64\times 64$ pixels; and each postage stamp contains at least one galaxy. These $10^4$ postage stamps, within one subfield, are divided into $5\times10^3$ orthogonal pairs. For two stamps within each pair, the orientations of the intrinsic galaxy images are 90 degree separated from each other to reduce the shape noise. Galaxies in each subfield are distorted by the same shear and subsequently smeared by the same PSF. Shear and PSF vary for different subfields. 

One difference between GREAT3-HSC simulations and the HSC-like BDK simulation is that GREAT3-HSC simulations use the HSC-like correlated noise model and PSF models. The noise of GREAT3-HSC simulations is generated with the autocorrelation function measured from the blank pixels of HSC coadd exposures \citep[see][fig.1]{GREAT3HSC17}. GREAT3-HSC simulations also use realistic PSF models of coadd HSC exposures which are constructed by the HSC pipeline. Whereas the HSC-like BDK simulation uses uncorrelated Gaussian noise and Moffat PSF. 
\begin{figure*}
\centering
\includegraphics[width=0.8\textwidth]{\figPath 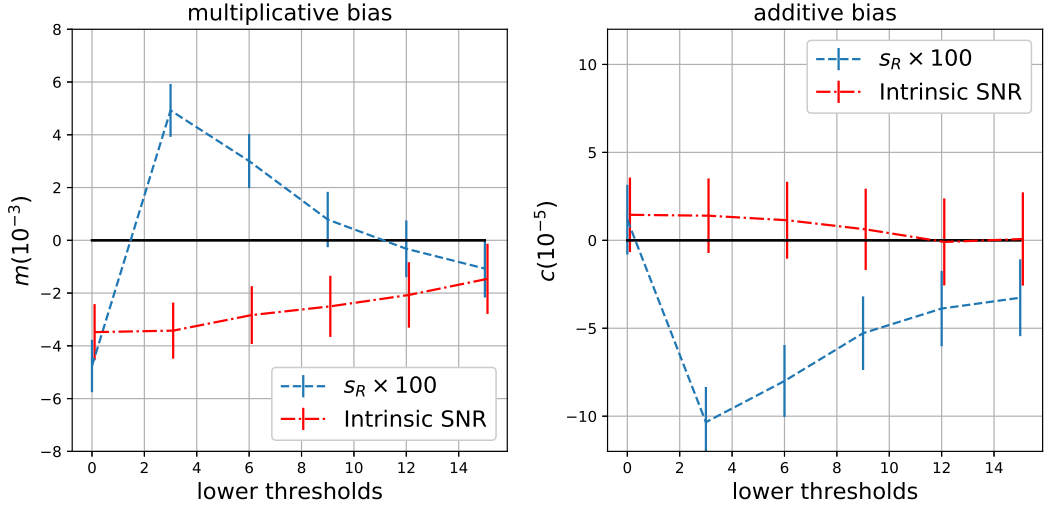}
\caption{The performance of the FPFS shear estimator on the HSC-like BDK simulation. (i) {The left panel} shows multiplicative bias as function of lower selection threshold. (ii) {The right panel} shows the additive bias as function of lower selection threshold. The red dotted-dashed lines show the result for intrinsic $S/N$ and the blue dashed lines show the result for FPFS flux revised by the iterative method proposed in Section \ref{sub:selectMeth}. The red dotted-dashed lines focus on noise bias and model bias, whereas the blue dashed lines also demonstrate the performance of selection bias revision. The free parameters are kept to the default values ($\alpha=4$, $\beta=0.85$, $\nu=4$) for these results. }
\label{fig_BDK}
\end{figure*}

Another difference is that the GREAT3-HSC simulations contain four galaxy samples with different galaxy properties. The galaxy properties for these four galaxy samples are described as follows. Sample 1 and sample 2 only contain one isolated galaxy on each postage stamp. Sample 1 is composed of realistic galaxies from the 25.2 magnitude limited COSMOS HST galaxy sample. Galaxies in sample 2 are parametric Sersic models which are fitted to the galaxies in the $25.2$ magnitude limited COSMOS HST sample. Sample 2 is similar with the HSC-like BDK simulation but the HSC-like BDK simulation simulates large number of galaxies ($4\times 10^{8}$ galaxies) without using orthogonal pairs to reduce the shape noise. 

Multiple galaxies can be found on each postage stamp of sample 3 and sample 4. Neither deblender nor noise replacer has been ran before inserting the HST galaxies into the postage stamps of the simulations. Therefore, neighbouring objects from the COSMOS HST survey are also included in sample 3 and sample 4. Sample 3 uses the postage stamps of the COSMOS HST survey and it applies the selection criteria of the COSMOS HST survey. Since the size of the HST postage stamps are generally smaller than the size of the postage stamps of the GREAT3-HSC simulations, there are no objects near the edge of the simulated stamps. Sample 4 does not truncate input galaxy images with the COSMOS HST stamps, so the COSMOS HST images extend to the edges of the simulated postage stamps and the HST images are artificially truncated by the edges of these postage stamps. Moreover, the density of footprints and the selection criteria for sample 4 are matched to the HSC observations. The summary of these four samples can be found in \citet[Table 1]{GREAT3HSC17}.

\subsection{Pipeline}
\label{sub:pipeline}
Our FPFS method is implemented into the HSC pipeline \citep{HSC1-pipeline}, which is an open-source software developed to process data observed by the ongoing HSC survey and the future LSST survey. The HSC pipeline performs a maximum likelihood analysis to detect pixels with a $5\sigma$ threshold from the simulated exposures. Every detected peak is defined as a source object and the connected nearby region above the threshold is identified as footprint of the source object. For a stamp which does not contain any detected footprint in the central region ($10\rm {pix} \times 10\rm {pix}$ around the center of the stamp), we assign a peak to the postage stamp center and a $10 \rm pix \times 10 \rm pix$ footprint around its peak. Such assigned source object is labeled as undetected. We use the HSC pipeline to subtract background from exposures in sample 3 and sample 4 to reduce light remnant from neighbouring objects. If a footprint contains multiple number of peaks, we use the HSC deblender to apportions the flux to different peaks. The HSC pipeline adopts the SDSS deblending algorithm \citep{SDSSpipe} as the first generation of deblender. It takes each peak as a `child' source of the `parent' source. With the assumption that every object has a 180-degree rotational symmetry around the peak, a template $T_i(\vec{x})$ for each `child' is defined as follows
\begin{align}\label{deblend_templete}
T_{i}(\vec{x})=\text{min} (f(\vec{x}),f(2\vec{p}_i-\vec{x})),
\end{align} 
where $\vec{p}_i$ is the peak of the `child' source $i$, $\vec{x}$ and $2\vec{p}_i-\vec{x}$ are symmetric about the peak $\vec{p}_i$. Then scaling parameters $c_i$ are deduced by fitting the templates to the `parent' image. The final deblended `child' source is
\begin{align*}
f^{D}_i(\vec{x})=\frac{c_i T_i(\vec{x})}{\sum_j c_j T_j(\vec{x})} f(\vec{x}).
\end{align*}
For each detected `child' source object, the HSC pipeline replaces the footprints of other sources with uncorrelated Gaussian noise. Shape of every source object is measured after the noise replacement. 

At the beginning of the FPFS shape measurement, we define the boundary of each galaxy using top-hat aperture defined in eq. (\ref{top-hat definition}). The center of the aperture is set to the position of the peak if the footprint contains only one peak. On the other hand, for a footprint with multiple peaks, the center of the aperture is set to be the footprint center. The aperture radius is set to $\alpha$ times of the half light radius of each galaxy , where the default $\alpha$ is determined in Section \ref{sub:cutRad} and the half light radius is the trace radius determined by re-Gaussianization algorithm. Furthermore, in order to ensure that the aperture region covers the entire footprint area for each galaxy, the minimum value of aperture radius is set to $r_{fp}+3$, where $r_{fp}$ is the radius of the footprint determined by the HSC pipeline. We then put each galaxy into a $64\times64$ stamp and pad the region outside aperture with zero. After Fast Fourier Transform (FFT), we calculate the galaxy Fourier power function and subtract the noise Fourier power function from it. Finally, the FPFS ellipticity and responsivity are calculated after deconvolving the PSF Fourier power function from the galaxy Fourier power function. 

{\blueColor
Our FPFS algorithm is written in python, using the public library numpy \citep{numpy2007}, and it is successfully implemented into HSC pipeline. The pipeline takes approximately $0.07$ second to conduct shape measurement on one galaxy. The code is released on GitHub \footnote{\url{https://github.com/superonion1993/FPFS}}.}
\begin{figure*}%[!ht]
\centering
\includegraphics[width=0.9\textwidth]{\figPath 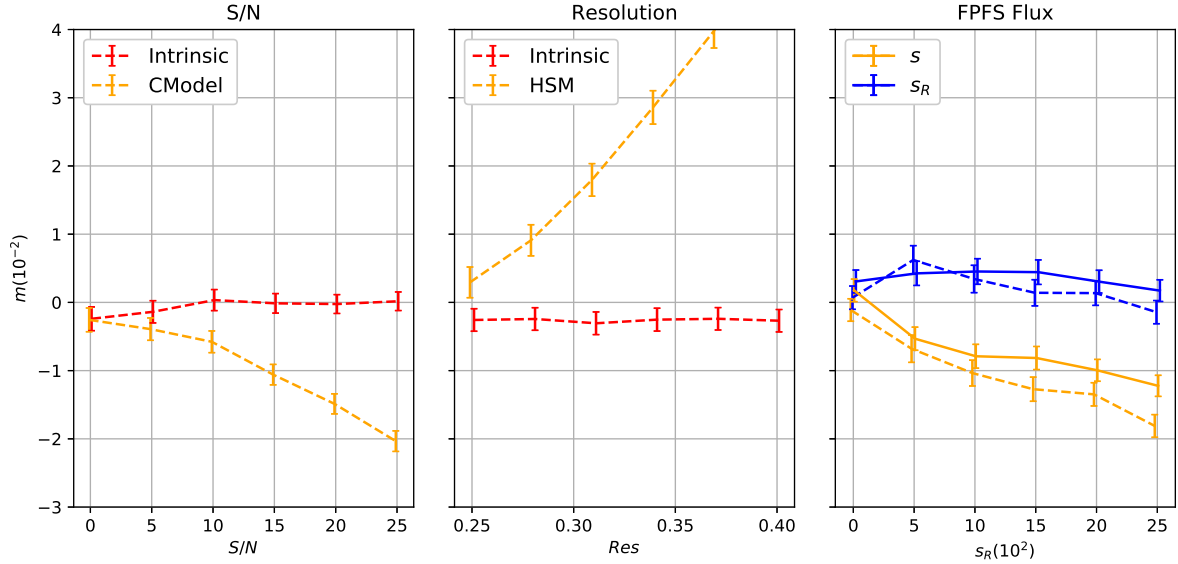}
\caption{The performance of the FPFS method on isolated galaxies (sample 1 and sample 2 of GREAT3-HSC simulations) with different selection functions and thresholds. (i) The left panel compares the selection of intrinsic $S/N$ (red line) and CModel $S/N$ (yellow line). (ii) The middle panel compares the selection with intrinsic resolution (red line) and resolution measured by the HSM algorithm (yellow line). (iii) The right panel compares the selection with the observed FPFS flux (yellow lines) and the revised FPFS flux (blue lines). The $x$-axis is the lower threshold and the $y$-axis represents the multiplicative bias. The solid lines represent tests done on sample 1 and the dashed lines represent tests on sample 2. The free parameters are kept to the default values ($\alpha=4$, $\beta=0.85$, $\nu=4$) for these results.}\label{fig_selectBias}
\end{figure*}

In the following subsection, shear will be measured from different simulations using the aforementioned pipeline. Subsequently, we conduct a linear fit of the measured shear ($\hat{g}_{1,2}$) by the input shear ($g_{1,2}$) to determine the multiplicative bias and additive bias defined in eq. (\ref{bias definition}). Furthermore, the error of the bias is determined by the covariance matrix of the fitting.

\subsection{Free parameters }
\label{sub:cutRad}

The accuracy and precision of the FPFS estimator is dependent on the setups of three free parameters ($\alpha$,$\beta$,$\nu$). As discussed in Section \ref{sec:Method}, $\alpha$ and $\beta$ determine the measurement scale in real space and Fourier space, respectively. Whereas $\nu$ changes the relative weight between galaxies with different luminosities and reduces noise bias.
\subsubsection{parameters $\alpha \& \beta$}
Firstly, we discuss the dependence of the accuracy on the values of ($\alpha$, $\beta$). LZ17 has studied how $\alpha$ influences the accuracy of the shear estimator proposed by ZK11 which corresponds to the case of $\nu=+\infty $ for the FPFS estimator. LZ17 kept $\beta=0.5$ as a constant so LZ17 did not consider the dependence of the accuracy on $\beta$. We notice that the accuracy of our FPFS estimator is dependent on both $\alpha$ and $\beta$. The multiplicative bias as function of ($\alpha$, $\beta$) is shown in Fig. \ref{fig_betaCut}. The additive bias is not plotted since it is only a few parts in $10^4$.  This test is conducted on sample 2 of GREAT3-HSC simulations and keeps $\nu$ to $4$ and selects the galaxies with intrinsic $S/N$ greater than $5$. \footnote{The intrinsic $S/N$ and intrinsic resolution are measured before the weak lensing distortions.} Fig. \ref{fig_betaCut} shows that, for $\beta=0.5$, the multiplicative bias exceeds one percent when the aperture ratio drops below $8$. This result is consistent with what we found in LZ17. Furthermore, we find that when $\beta$ is increased to $0.85$, the multiplicative bias is consistent with zero even if $\alpha$ drops to $4$. We conclude that the region extended to four times of each galaxy's half light radius is required for the accurate shear measurement if $\alpha$ is kept to $0.85$. Therefore, we set $\beta=0.85$ and $\alpha=4$ as the default parameters in the following content. Fig. \ref{fig_galCut} shows the galaxy images and the corresponding mask image detected from sample 2 of GREAT3-HSC simulations. The connected white pixels on the mask images are the footprints of galaxies and the circular gray pixels on the mask image are the aperture.

\subsubsection{parameter $\nu$}
Secondly, we discuss the dependence of the accuracy on the value of $\nu$. We change the value of $\nu$ and conduct shear measurement on sample 2 of GREAT3-HSC simulations. The multiplicative bias as a function of $\nu$ is shown in Fig. \ref{fig_AccuracyNu}. The additive bias is not plotted since it is only a few parts in $10^4$. In summary, when $\nu$ increases beyond $1$, the multiplicative bias is reduced below one percent.

{\redColor 
Finally, we discuss the dependence of the precision on the value of $\nu$. $\nu$ influences the precision of shear estimation by adjusting the weight between galaxies with different luminosities. To demonstrate, we separate galaxies in sample 2 of GREAT3-HSC simulations into different bins according to their intrinsic $S/N$ and plot the averaged $R_1$ in each bin with different values of $\nu$ in the left panel of Fig. \ref{fig_PrecisionNu}. The case where weight ratio is proportional to $S/N$ is also plotted in the left panel of Fig. \ref{fig_PrecisionNu} as a reference.

Measurement error is generally used to quantify the precision of shear estimation. We study the influence of $\nu$ on the measurement error caused by the shape noise and the photon noises using $8\times10^4$ modelled galaxies randomly selected from the $25.2$ magnitude limited COSMOS HST catalog (without repeated selection). The setups of the simulation is described as follows.
\begin{enumerate}
\item Rotate the $8\times10^4$ galaxies with different random angles.
\item Convolve each sheared galaxy with a Moffat PSF ($\beta_{\rm{psf}}=3.5$, $e_1=0.00$, $e_2=0.025$).
\item Add random photon noises to the galaxies with HSC noise level.
\item Measure shear from the galaxy ensemble with different choice of $\nu$.
\end{enumerate}
We repeat step (i) to step (iv) with different random seeds for orientations and photon noises. Shear is repeatedly measured from each realization with different choices of $\nu$ and different selection thresholds on $M_{00}$. According to the central limit theorem, the measurement error for these realizations follow a Gaussian distribution, therefore the measurement error can be quantified by the standard deviation of the measured shear. We rescale the measurement error by $\sqrt{N}$ to calculate the shear measurement error per galaxy, where $N$ is the total number of galaxies used for shear measurement. 

The measurement errors per galaxy per shear component for different $\nu$ and different lower selection thresholds on $M_{00}$ are plotted in the middle panel of Fig. \ref{fig_PrecisionNu}. The right panel of Fig. \ref{fig_PrecisionNu} shows the galaxy number for different lower selection thresholds. As shown in the middle panel of Fig. \ref{fig_PrecisionNu}, there exists an optimal setup of $\nu$, within the range between $\nu=1$ and $\nu=2$ , for each selection criteria. The optimal $\nu$ slightly overweight bright galaxies since the errors on faint galaxies are generally larger than the errors on bright galaxies since the bright galaxies have greater Signal to Noise Ratio (S/N).

{\blueColor
We also compare the measurement error between FPFS and re-Gaussianization using galaxy images from the first year HSC survey. We randomly rotate the calibrated ellipticities measured by re-Gaussianization and FPFS. The measurement errors for two methods are subsequently determined. In the procedure of determining the measurement error, the uncertainty of shear calibration is assumed to be zero. The details of this experiment are shown in Appendix \ref{app:compareReG}. The measurement error for FPFS, when setting $\nu=4$, is about $23\%$ larger than that of re-Gaussianization. Although the results in Appendix \ref{app:compareReG} shows that re-Gaussianization has lower noise level, re-Gaussianization needs external galaxy image simulations to calibrate noise bias and model bias even for isolated galaxies. However, the uncertainty of such calibration has not been taken into account. Finally, we conclude that the FPFS method has to increase the noise level by about $23\%$ to reduce noise bias below $1\%$. As shown in \citet{metacal2}, Metacalibration adds artificial noises, which are inversely sheared, to galaxy images to reduce noise bias. Such procedure of Metacalibration also increases measurement error by about $20\%$ \citep[see section 11 of][]{metacal2}.}

In the presence of blending, shear estimator should also be calibrated for blending bias due to the imperfection of deblending algorithms. As shown in Section \ref{sub:neiObj}, blending bias for faint galaxies is much larger than that for bright galaxies. Therefore, bright galaxies are more reliable than faint galaxies. Consequently, we set $\nu=4.0$ as the default parameter to add more weight to bright galaxies. Moreover, such setup also ensures that noise bias is properly reduced.
}

\begin{figure*}%[!ht]
\begin{center}
\includegraphics[width=0.8\textwidth]{\figPath 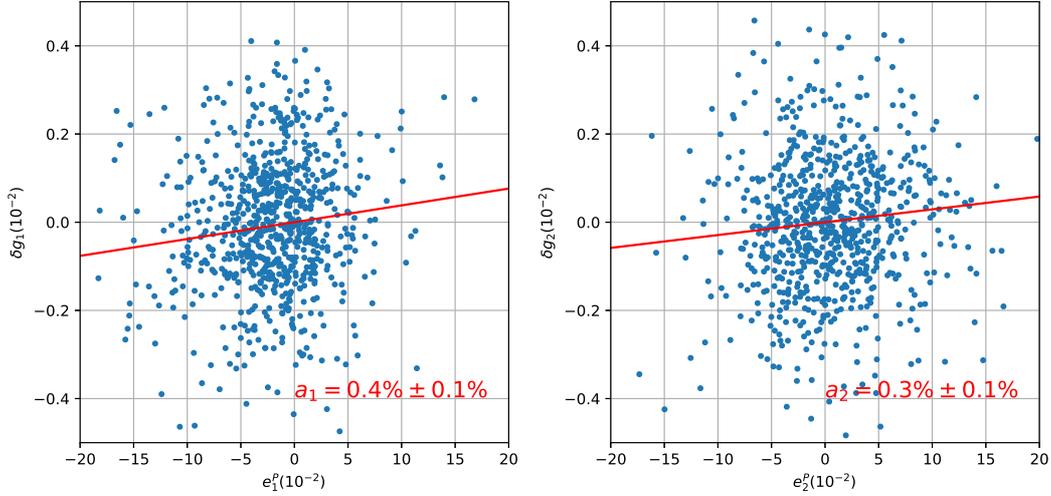}
\end{center}
\caption{The dependence of the additive bias on the PSF anisotropy. The test is conducted on sample 2 of GREAT3-HSC simulations. The left (right) panel shows the relation between the shear residual of $g_1$ ($g_2$) and PSF ellipticity $e^P_1$ ($e^P_2$). Each blue point is the result for one subfield. The red lines in two pannel are the fitting relations for the two components of the shear residual and PSF ellipticity, respectively. The x-axises represent the PSF ellipticity and the y-axis represent the shear residual. The free parameters are kept to the default values ($\alpha=4$, $\beta=0.85$, $\nu=4$) for these results.}\label{fig_psfAniso}
\end{figure*}

\subsection{Isolated galaxies}
\label{sub:selectTest}
\subsubsection{HSC-like BDK}
\label{subsub:BDK}
We firstly run our pipeline on the HSC-like BDK simulation. Subsamples of galaxies are selected from the HSC-like BDK simulation with different selection functions including the intrinsic $S/N$ measured before lensing and the FPFS flux revised using the iterative method proposed in Section \ref{sub:selectMeth}. Shear is measured with the default parameters ($\alpha=4$, $\beta=0.85$, $\nu=4$). The results for different lower selection thresholds are shown in Fig. \ref{fig_BDK}.

The red lines in Fig. \ref{fig_BDK} show the results selecting galaxies with intrinsic $S/N$. These results do not include selection bias by construction and only focus on noise bias and model bias for isolated galaxies. For samples selected with the intrinsic observables, the amplitude of multiplicative bias is well below $5 \times 10^{-3}$ and the amplitude of additive bias is below $5 \times 10^{-5}$. 

The blue lines in Fig. \ref{fig_BDK} show the results selecting galaxies with the revised FPFS flux. These results demonstrate the performance of our iterative method which is developed to revise selection bias. The multiplicative bias fluctuates within $1\times 10^{-2}$ and additive bias fluctuates within $1\times 10^{-4}$. 

Selection bias is only reduced below $1\%$ since we have not consider the influence of noise in the transformation equation of FPFS flux shown in eq. (\ref{select_transform}). However, such remaining bias is below the first year science requirement of the HSC survey given by \citet{HSC1-catalog}.
\subsubsection{GREAT3-HSC}
\begin{table*}%[!ht]
\centering
\begin{tabular}{l*{3}{c}{r} }
\hline \hline
sample \& setup & $m_1(10^{-2})$& $c_1(10^{-4})$& $m_2(10^{-2})$ &$c_2(10^{-4})$  \\ \hline
S3ND &$-0.25 \pm 0.22$ & $0.75 \pm 0.56$ & $0.03 \pm 0.23$  & $-0.71\pm 0.59$  \\
S3D  &$-5.71 \pm 0.24$ & $3.33 \pm 0.60$ & $-5.59 \pm 0.24$ &$-1.06 \pm 0.60$ \\
S4ND &$-1.68 \pm 0.27$ & $0.24 \pm 0.71$ & $-1.11 \pm 0.23$ & $ 0.19\pm 0.57$  \\
S4D  &$-5.83 \pm 0.41$ & $1.27 \pm 1.06$ & $-5.59 \pm 0.30$ & $-0.71\pm 0.75$ \\
 \hline
\end{tabular}
\caption{Performance of the FPFS method on sample 3 and sample 4 with 2 different setups. The column `sample \& setup' shows the sample and the setup of the corresponding experiment. The free parameters are kept to $(\alpha=4,\beta=0.85,\nu=4)$ for these results. }\label{tab:S3S4Test}
\end{table*}
We subsequently test the FPFS method using galaxies in sample 1 and sample 2 of the GREAT3-HSC simulations. These galaxy samples only contain isolated galaxies. Shears are measured from these galaxy samples with different selection functions and different lower selection thresholds. These selection functions include intrinsic $S/N$, observed CModel $S/N$, intrinsic resolution, observed resolution, observed FPFS flux, and revised FPFS flux. The free parameters are set to the default values ($\alpha=4$, $\beta=0.85$, $\nu=4$). The multiplicative biases for these selections are demonstrated in Fig. \ref{fig_selectBias}.

As demonstrated in the left panel and the middle panel of Fig. \ref{fig_selectBias}, the multiplicative biases are blow $5 \times 10^{-3}$ if the intrinsic quantities (intrinsic $S/N$ or intrinsic $resolution$) are used to select galaxies. {\blueColor These results which select galaxies with the intrinsic observables do not include selection bias by construction.} 

On the other hand, if the observed quantities (observed CModel $S/N$ or observed resolution) are used to select galaxies, multiplicative biases grow as the lower selection thresholds grow. {\blueColor These results include not only noise bias and model bias but also selection bias.

The right panel of Fig. \ref{fig_selectBias} demonstrates the results selecting galaxies with observed FPFS flux and revised FPFS flux. The FPFS flux is revised using the iterative method introduced in Section \ref{sub:selectMeth}. As demonstrated in the right panel of Fig. \ref{fig_selectBias}, the iterative method reduces selection bias below $1\%$. These results are in consistent with what we have found in Section \ref{subsub:BDK}}.

Furthermore, the results for sample 1 and sample 2, as shown in the right panel of Fig. \ref{fig_selectBias}, are consistent with each other even though the galaxy morphology in these two samples are different. Therefore, we conclude that the FPFS shear estimation for isolated galaxies is not influenced by model bias.

Finally, we test the dependence of additive bias on PSF anisotropy. This test is conducted on sample 2 of the GREAT3-HSC simulations. Since sample 1 and sample 2 use the same PSF models, we do not repeat this test on sample 1. The ellipticity of PSF, measured by the re-Gaussianization algorithm, is used to quantify PSF anisotropy. Setting the fiducial multiplicative bias to zero, we fit the PSF ellipticity to the shear residual according to the linear relationship
\begin{equation}
\delta g_{1,2}=\hat{g}_{1,2}-g_{1,2}=a_{1,2} e^{P}_{1,2},
\end{equation}
where $\delta g_{1,2}$ is the shear residual, $\hat{g}_{1,2}$ is the estimated shear, $g_{1,2}$ is the input shear, $e^{P}_{1,2}$ is the PSF ellipticity and $a_{1,2}$ is termed fractional additive bias which describes the fraction of the PSF anisotropy which leaks into the shear measurement \citep{GREAT3HSC17}.

Fig. \ref{fig_psfAniso} demonstrates the relation between the shear residual and the PSF ellipticity. We mask out the data point with extreme ellipticity, where the amplitude of the ellipticity is greater than $0.3$. These masked data only corresponds to $0.4\%$ of the total data. As demonstrated in Fig. \ref{fig_psfAniso}, the fractional additive bias is only $(4 \pm 1) \times 10^{-3}$. This result not only confirms that the additive bias is far below the first year HSC science requirements, it also proves that the additive bias is almost not correlated with the PSF anisotropy.

However, we caution that GREAT3-HSC simulations do not include PSF model residuals due to the uncertainty in the PSF reconstruction. We leave the systematic tests including the PSF model residuals to our future work.  

\subsection{Blended galaxies}
\label{sub:neiObj}
In this subsection, we conduct tests on sample 3 and sample 4. These samples also contain {\blueColor blended} galaxies so they are more close to the real observations.  We process all of the detected source objects in these samples with two different strategies, namely `Deblended' and `Nondeblended'. For the `Deblended' cases, we run the HSC deblender if multiple peaks are detected on a footprint before shape measurement. On the other hand, for the `Nondeblended' cases, we do not deblend any footprint even though multiple peaks exist in one footprint. 
`S3D' and `S3ND' represent `Deblended' and `Nondeblended' cases for sample 3. `S4D' and `S4ND' represent `Deblended' and `Nondeblended' cases for sample 4. We use the default parameters ($\alpha=4$, $\beta=0.85$, $\nu=4$) and select galaxies with the criterion: $s_R>1.5\%$.

The results of these tests are laid out in Table \ref{tab:S3S4Test}. Since the additive bias is below $4\times 10^{-4}$, we focus our discussion on the multiplicative bias. Comparing the results of `S3D' and `S4D' with the results on isolated galaxies shown in Section \ref{sub:selectTest}, we conclude that the HSC deblender fails to recover the true galaxy light profiles precisely and the discrepancy between the deblended galaxies and the true galaxies causes the multiplicative bias. The possible origins of the discrepancy are listed as follows.
\begin{enumerate}
\item The HSC deblender assumes a 180-degree rotational symmetry around the peak, although galaxies could have some irregular shape.
\item The HSC deblender tends to use the pixels, where the errors caused by photon noises are negative, to construct template.
\item The HSC deblender changes the autocorrelation function of photon noises within the source footprint.
\end{enumerate}

The multiplicative bias for `S3ND' is below $1\%$. However, we find a multiplicative bias about $1.5\%$ for `S4ND'. Considering the difference between the observational conditions of sample 3 and sample 4 as shown in Section \ref{sub:pipeline}, such small bias for `S4ND' must be caused by the contamination of light from the neighbouring objects.
The results of `S3ND' and `S4ND' indicate that it is possible to directly measure a `parent' source without deblending, if all of its `child' sources are on the same redshift plane and distorted by the same shear signal. The FPFS shear estimator works well on such `parent' sources since the FPFS algorithm is not sensitive to the off-centering effect and do not make any assumption on galaxy morphology. However, it is difficult to apply such strategy to real observations since the `child' galaxies of a `parent' source could be located on different redshift planes. 

\begin{figure*}%[!htb]
\begin{minipage}[!t]{.48\textwidth}
\includegraphics[width=1.\textwidth]{\figPath 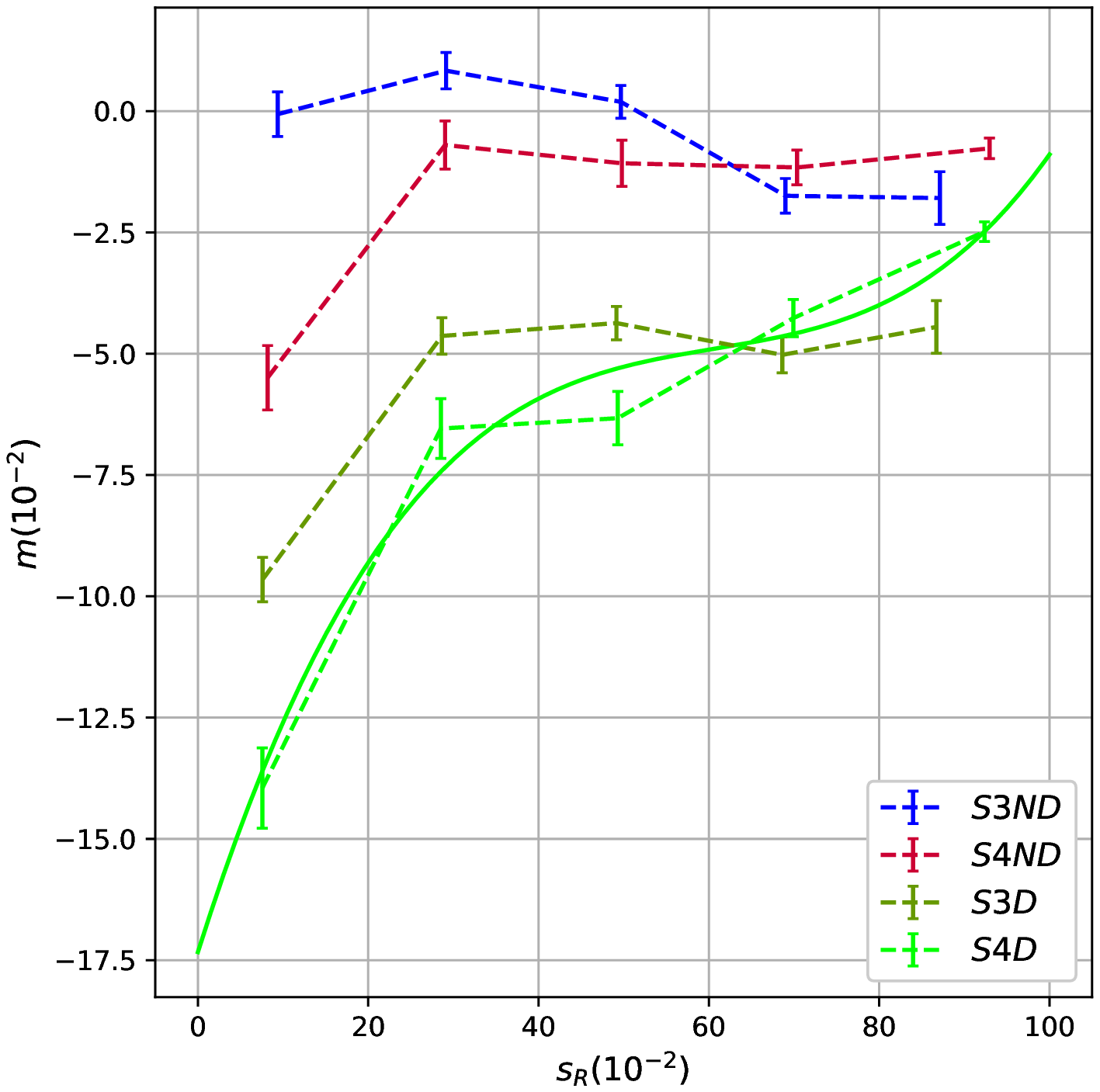}
\caption{The multiplicative bias for $s_R$ bins for sample 3 and sample 4. Lines labelled with `S3ND' and `S3D' are `Deblended' and `Nondeblended' cases of sample 3, while `S4ND' and `S4D' represent the two cases of sample 4. The $x$-axis is the revised FPFS flux ($s_R$) bins and the $y$-axis is the multiplicative bias ($m$). The solid lines are the third order polynomial fittings of $m(s_{R})$. The free parameters are kept to the default values ($\alpha=4$, $\beta=0.85$, $\nu=4$) for these results.}\label{fig_biasCalib}
\end{minipage} \hfill
\begin{minipage}[!t]{.48\textwidth}
\centering
\includegraphics[width=1.\textwidth]{\figPath 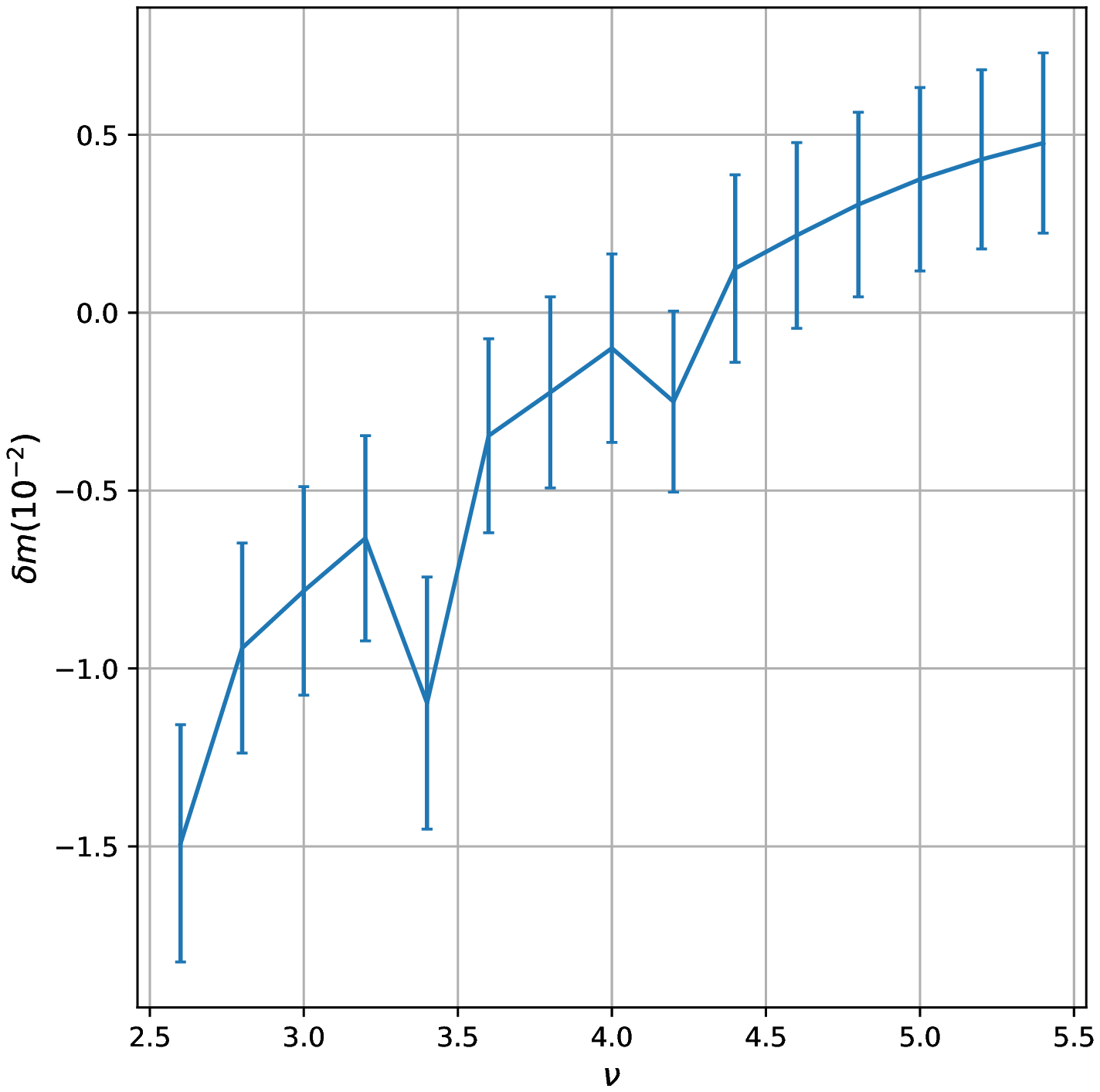}
\caption{The stability of the calibration factor under the distortion on $\nu$. The calibration factor obtained from the default setup ($\nu=4$) is tested under conditions whose setups deviate from the default. The $x$-axis shows the setups of $\nu$ in the consistency test and the $y$-axis is the uncertainty of multiplicative bias. The aperture ratio and shapelets scale ratio are kept to ($\alpha=4$,$\beta=0.85$) for these results. }\label{fig_changeNu}
\end{minipage}
\end{figure*}
With the intent to further understand the blending bias, we separate galaxies into different $s_R$ bins and determine the multiplicative bias using the galaxies within each bin. The multiplicative bias for each $s_R$ bin is demonstrated in Fig. \ref{fig_biasCalib}. For `S3ND' and `S4ND' cases, the multiplicative biases converge to approximately $-1.8\%$ ($6\sigma$ significance) at the bright end. The default aperture ratio ($\alpha$) determined in Section 3.2 would be too small to accurately measure the shear from the `parent' sources containing multiple number of bright objects. Since we are not going to apply the `Nondeblended' strategy to the real observation, the proper boundaries for blended `parent' sources are not discussed in this paper.

Another line of thought is to calibrate the blending bias with the GREAT3-HSC simulations. We focus on sample 4 since it best matches the real observational conditions. The results of `S4D' is consistent with our expectation that the deblender performs better on bright galaxies than on faint galaxies. As demonstrated by the green line in Fig. \ref{fig_biasCalib}, we conduct a third order polynomial fitting of the multiplicative bias as a function of the revised FPFS flux.

After modeling the multiplicative bias as a function of the revised FPFS flux. The calibration factor $1+m(s_R)$ is added to the responsivity $R_{1,2}$ to calibrate blending bias. It is worth noting that substituting the revised FPFS flux $s_R$ with the observed FPFS flux $s$ does not change the expectation value of calibrated responsivity, since
\begin{align*}
\left\langle m(s)R_{1,2}\right\rangle=\left\langle m(s_R)R_{1,2}\right\rangle.
\end{align*}
This equation is deduced by substituting eq. (\ref{select_transform}) into $\left\langle m(s_R)R_{1,2}\right\rangle$. The calibrated shear estimator is finally constructed as
\begin{align}\label{estimator_revised}
\hat{g}_{1,2}=\frac{\left\langle e_{1,2}\right\rangle}{\left\langle(1+m(s))R_{1,2}\right\rangle}.
\end{align}

\section{Consistency Test}
We have shown that the systematic biases on the FPFS estimator are well controlled below $1\times 10^{-2}$ for isolated galaxies in the previous section. However, the multiplicative bias originating from blending has to be calibrated by the GREAT3-HSC simulations. In this section, we check the consistency and stability of the calibrated estimator defined in eq. (\ref{estimator_revised}). 

We start off by using this revised shear estimator with the default setup of parameters to measure shear from sample 4. Galaxies are selected with $s_R>1.5\%$. Table \ref{tab:consisFull} demonstrates two results, where the calibration factor is constructed by $s$ or $s_R$. The remaining biases are labelled as $\delta m$ and $\delta c$. As shown in Table \ref{tab:consisFull}, the remaining biases are consistent with zero. By construction no systematic biases are found in these consistency tests since the calibration factors are deduced from the simulations with exactly the same galaxy sample and exactly the same setup of parameters. 

However, it is reasonable to investigate the performance of the calibrated estimator under the following conditions. (i) The setups of the FPFS method, in real observation, deviate from the default setup. (ii) The observational conditions of the data, to which the calibrated estimator is applied, are different from that of sample 4.
\begin{table}%[!h]
\centering
\begin{tabular}{l*{2}{c}r}
\hline \hline
sample&calib	 & $\delta m(10^{-2})$     & $\delta c(10^{-4})$\\ \hline
full&$s_R$       & $-0.09 \pm 0.27$ & $0.31 \pm 0.69$ \\
full&$s$         & $-0.10 \pm 0.26$ & $0.29 \pm 0.68$ \\ \hline
\end{tabular}
\caption{The performance of the calibration on the full sample (sample 4 of GREAT3-HSC simulations). $s_R$ and $s$ represent the result where calibration factors are constructed with the revised FPFS flux and the observed FPFS flux, respectively. The free parameters are kept to $(\alpha=4,\beta=0.85,\nu=4)$ for these results.}\label{tab:consisFull}
\end{table}

With the intent to check the stability of the calibrated estimator under the distortion of the setup, we change $\nu$ to be different from the default setup ($\nu=4$) and test the performance of the calibration factor obtained from the default setup. The remaining multiplicative bias ($\delta m$) is shown in Fig. \ref{fig_changeNu}. The remaining additive bias ($\delta c$) is not plotted since it is only a few parts in $10^{5}$, which is far below the first year science requirement of HSC survey \citep{HSC1-catalog}. From Fig. \ref{fig_changeNu}, we find that $\delta m$ is approximately proportional to the distortion around $\nu=4$, which is our default value. Even though the distortion of $\nu$ increases to $40\%$ of the default value, the amplitude of the remaining bias is below $1.5\%$. 

In order to check the stability of the calibrated estimator under the distortion of observational conditions, we separate galaxies into four quartiles based on PSF FWHM in the same way as \citet{GREAT3HSC17} and use the calibrated estimator to measure shear within each quartile. The result is shown in Table \ref{tab:validity}. The uncertainty of multiplicative bias ($\delta m$) is below $1\times 10^{-2}$ and the uncertainty of additive bias ($\delta c$) is below $2.5$ part in $10^4$.

We caution that the calibration factor derived from sample 4 of the Great3-HSC simulations can only be applied to samples with similar blending conditions. For example, if we apply the calibration factor derived from sample 4 to the measurements conducted on galaxies in sample 1, the final shear estimation will be biased. Since sample 1 only contains isolated galaxies, whereas sample 4 contains both blended and isolated galaxies. The blending conditions for these samples are different. 
\begin{table}%[!h]
\centering
\begin{tabular}{l*{2}{c}r}
\hline \hline
sample&calib	 & $\delta m(10^{-2})$     & $\delta c(10^{-4})$\\ \hline
1st quartile&$s$ & $0.88  \pm 0.56$ & $-2.15\pm 1.48$ \\ 
2nd quartile&$s$ & $-0.36 \pm 0.30$ & $2.03 \pm 0.73$ \\ 
3rd quartile&$s$ & $-0.49 \pm 0.54$ & $1.83 \pm 1.35$ \\ 
4th quartile&$s$ & $-0.70 \pm 0.86$ & $-1.33\pm 2.23$ \\ \hline
\end{tabular}
\caption{The performance of the calibrated estimator on subsamples. Sample 4 of the GREAT3-HSC simulations are divided into 4 quartiles with different PSF properties. The free parameters are kept to $(\alpha=4,\beta=0.85,\nu=4)$ for these results.}\label{tab:validity}
\end{table}
\section{Summary and Outlook}
The newly developed FPFS method is based on the mathematical foundations of ZK11 and shapelets. The FPFS method projects galaxy's Fourier power functions onto four shapelet basis vectors after PSF deconvolution in Fourier space. Using four shapelet modes (Fig. \ref{fig_MnmHist}) measured from each galaxy, ellipticity and responsivity (Fig. \ref{fig_ElliHist}) are constructed. The transformation formula of ellipticity under the influence of shear for each single galaxy is derived. Based on the transformation formula, the shear estimator is finally given by eq. (\ref{estimator_define}). 

The FPFS formalism introduces several free parameters ($\alpha$,$\beta$,$\nu$). $\alpha$ and $\beta$ determine the measurement scale in real space and Fourier space, respectively. The influence of ($\alpha$, $\beta$) on the accuracy of shear estimation is shown in Fig. \ref{fig_betaCut}. Noise bias is reduced well below one percent by increasing the free parameter $\nu$ (Fig. \ref{fig_AccuracyNu}). Furthermore, $\nu$ changes the relative weight on different galaxies (left panel of Fig. \ref{fig_PrecisionNu}) and influence the precision of shear estimator (middle panel of Fig. \ref{fig_PrecisionNu}). We define the FPFS flux in eq. (\ref{select_define}). Based on the transformation formula of the FPFS flux, we propose an iterative method to reduce selection bias below one percent (right panel of Fig. \ref{fig_selectBias}).
Using the default setup ($\alpha=4$, $\beta=0.85$, $\nu=4$), our estimator is tested on several HSC-like image simulations. The results show that, for isolated galaxies, the amplitude of multiplicative bias is well below $1 \times 10^{-2}$. However, the uncertainty caused by the shape noise is increased by about $23\%$ comparing with the ideal optimal estimator.

We also test the FPFS shear estimator on more realistic samples which also contain blended galaxies. The blended galaxies are deblended by the first generation HSC deblender and multiplicative bias of $(-5.71\pm 0.31) \times 10^{-2}$ is found. (Table \ref{tab:S3S4Test}). Using sample 4 of the simulations, we model blending bias as a function of the FPFS flux and then calibrate it on single galaxy level (Fig. \ref{fig_biasCalib}). Several consistency tests for the calibration are conducted, which include distorting the default parameter (Fig. \ref{fig_changeNu}) and applying the calibration to subsamples of sample 4 (Table \ref{tab:validity}). We report that no significant remaining bias has been found. 

The following problems have not been discussed within this work.
\begin{enumerate}
\item The revision for selection bias is not perfect and the remanent multiplicative bias fluctuate within one percent.
\item Calibrations which are more stable and universal remain to be discovered.
\item The influence of PSF model residuals on shear estimation has not been discussed.
\item The influence of astrometry errors on shear estimation has not been discussed.
\item The performance of background subtraction and its influence on shear estimation has not been discussed.
\item {\redColor The difference between noise correlation on blank pixels and pixels with detected object on coadd exposures due to Poisson noise has not been discussed.}
\end{enumerate} 
These possible biases will be quantified using more realistic simulations in our future work.

\section*{Acknowledgements}
We thank Rachel Mandelbaum and Hironao Miyatake for making GREAT3-HSC simulations available, Naoki Yasuda for the instruction on the HSC pipeline, Ryoma Murata, Naoki Yoshida, Masahiro Takada, Robert Armstrong and Shuai Zha for useful discussions, Jun Zhang and Rachel Mandelbaum for useful comments and Jennifer Lau for proofreading.

This work was supported by Global Science Graduate Course (GSGC) program of University of Tokyo and JSPS KAKENHI (JP15H05892, JP18K03693).

\bibliography{\bibPath weak_lensing,\bibPath other}

\appendix
\section{Transformation of FPFS ellipticity}
\label{app:elliTrans}
We derive the transformation formula of the FPFS ellipticity in the absence of noise in this appendix. Starting from the first component of the FPFS ellipticity, we substitute eq. (\ref{Shapelets_Moments_shear_Transform}) into eq. (\ref{ellipticity_define}) and obtain
\begin{align*}
e_1&=\frac{M_{22c}}{M_{00}+C}\\
&=\frac{\bar{M}_{22c}-\frac{\sqrt{2}}{2}g_1(\bar{M}_{00}-\bar{M}_{40})+\sqrt{3}g_1\bar{M}_{44c}+\sqrt{3}g_2\bar{M}_{44s}}{\bar{M}_{00}+C+\sqrt{2}(g_1\bar{M}_{22c}+g_2\bar{M}_{22s})}.
\end{align*}
Since $g_{1,2}\ll 1$, the denominator can be expressed as the first order Taylor expansion of $g_{1,2}$ as
\begin{align*}
e_1&=\left(\frac{\bar{M}_{22c}-\frac{\sqrt{2}}{2}g_1(\bar{M}_{00}-\bar{M}_{40})+\sqrt{3}g_1\bar{M}_{44c}+\sqrt{3}g_2\bar{M}_{44s}}{\bar{M}_{00}+C}\right)\\
&\times \left(1-\sqrt{2}g_1\frac{\bar{M}_{22c}}{\bar{M}_{00}+C}-\sqrt{2}g_2\frac{\bar{M}_{22s}}{\bar{M}_{00}+C}\right).
\end{align*}
Subsequently, we neglect the terms which contain the second order of $g_{1,2}$ and obtain 
\begin{align*}
e_1= \bar{e}_1-g_1\bar{R}_1+\sqrt{3}g_1\frac{\bar{M}_{22c}}{\bar{M}_{00}+C}+\sqrt{3}g_2\frac{\bar{M}_{22s}}{\bar{M}_{00}+C}.
\end{align*}
Similarly, the transformation formula of $e_2$ under the influence of shear can be derived as
\begin{align*}
e_2= \bar{e}_2-g_2\bar{R}_2-\sqrt{3}g_2\frac{\bar{M}_{22c}}{\bar{M}_{00}+C}+\sqrt{3}g_1\frac{\bar{M}_{22s}}{\bar{M}_{00}+C}.
\end{align*}

\section{Fourier Power of Noise}
\label{app:noiPow}
In this appendix we discuss the reconstruction of the Fourier power function of noise from the autocorrelation of noises. The discrete form of noises on a $N\times N$ stamp is denoted as $h[\vec{n}]$, and $H_V[\vec{n}]$ and $H_R[\vec{n}]$ are used to denote the autoconvolution and the autocorrelation of noise. The pixels outside the aperture radius are masked with $0$. The definition of $H_V[\vec{n}]$ and $H_R[\vec{n}]$ are defined as follows:
\begin{equation}
\begin{split}
H_V[\vec{m}]&=\sum_{n_1=0}^N \sum_{n_2=0}^N h[\vec{n}]h[\vec{n}+\vec{m}],\\
H_R[\vec{m}]&=\sum_{n_1=0}^N \sum_{n_2=0}^N h[\vec{n}]h[\vec{n}+\vec{m}]/W[\vec{m}],
\end{split}
\end{equation}
where $W[\vec{m}]$ is the total number of pixel pairs separated by $\vec{m}$ in which both pixels are within the aperture.
Therefore, we can reconstruct autoconvolution of noise on the stamp by
\begin{equation}
H_V[\vec{m}]=H_R[\vec{m}] \times W[\vec{m}].
\end{equation}
The Fourier power function of galaxy can be obtained by doing the Inverse Fourier Transform (IFT) of the autoconvolution of noise \citep{Li17Auto}.

\section{Precision comparison with reGaussanization}
\label{app:compareReG}
\begin{figure}
\centering
\includegraphics[width=0.45 \textwidth]{\figPath 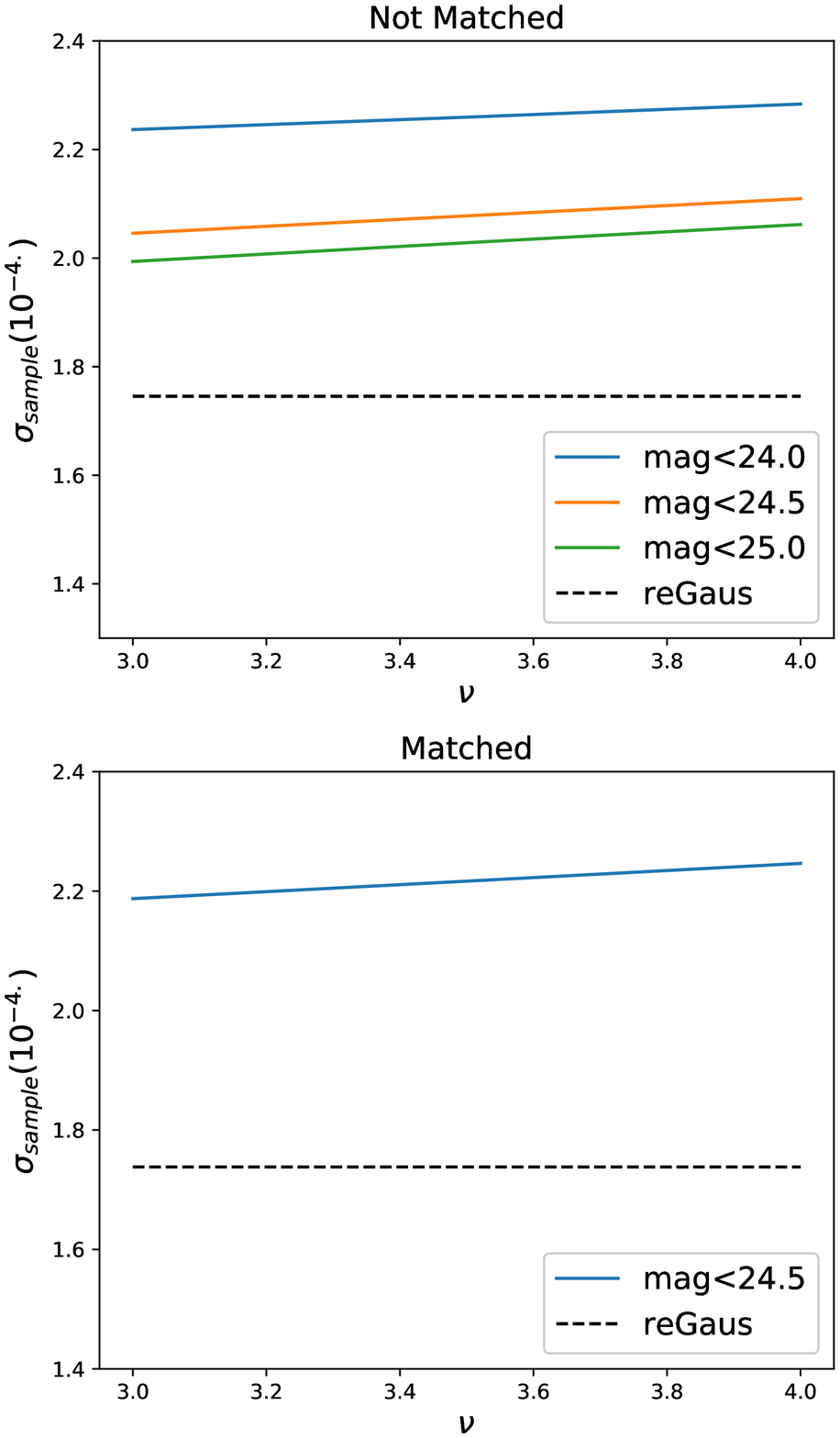}
\caption{Comparison of the precision between FPFS and re-Gaussianization using galaxies in GAMA15H field of the fist year HSC data release. (i) The top panel demonstrates the error of two methods with different selection criteria for different methods. (ii) The bottom panel demonstrates the error of two methods on the same galaxy sample.  The x-axises are the weighting parameter $\nu$. The y-axises are errors for the whole samples. Solid lines with different colors demonstrate the results with different magnitude cut for FPFS. The black dashed line is the result for re-Gaussianization. For FPFS method, the free parameters are kept to $(\alpha=4,\beta=0.85)$.}
\label{fig_compareRG}
\end{figure}
{\redColor
We use galaxy images in the GAMA15H wide field of the HSC first year data release \citep{HSC1-data} to compare the precision between FPFS and re-Gaussianization. Galaxy shapes are measured using both FPFS and re-Gaussianization. The measured ellipticities for each method are rotated by random angles. After repeating the rotation with different random seeds for $500$ times, we measure shear from these $500$ realizations. Calibration factors determined by GREAT3-HSC simulations are added to the estimators of both re-Gaussianization and FPFS. The standard deviations of these measurements are used to quantify the precision for two methods. The results are shown in Fig. \ref{fig_compareRG}. For re-Gaussianization method, we apply the standard weak lensing cut defined by \citet{HSC1-catalog}, whereas for FPFS method we apply the basic Full-Depth-Full-Color cut \citep{HSC1-catalog} and basic shape cuts: $M_{00}>500$, $|R_{1,2}|<5$. For the results shown in the top panel of Fig. \ref{fig_compareRG}, the galaxy samples for two methods are not exactly the same. The $24.5$ magnitude limited galaxy sample for FPFS are quite similar to the standard re-Gaussianization sample which also applies a $24.5$ magnitude cut. Whereas, for the results shown in the bottom panel of Fig. \ref{fig_compareRG}, we match between galaxy samples for two methods and force then to use the matched sample to determine the measurement error. Comparing the shear measurement error between FPFS and re-Gaussianization using the matched sample, we conclude that the error for FPFS are about $23\%$ larger than re-Gaussianization. The FPFS method reduces the multiplicative bias for isolated galaxies well below $1\times 10^{-2}$, however it increases measurement error by about $23\%$, which is a trade off between accuracy and precision. 
}

\bsp	% typesetting comment
\label{lastpage}
\end{document}